\documentclass[rmp,twocolumn]{revtex4-1}
\usepackage[a4paper,top=2.00cm,bottom=2.00cm,left=2.00cm,right=2.00cm]{geometry}
\usepackage{amsfonts}
\usepackage{amscd}
\usepackage{bm}
\usepackage{amsmath}
\usepackage{amssymb}
\usepackage{mathptmx}
\usepackage{color}
\usepackage{hyperref}
\usepackage{esint}
\usepackage{ifpdf}
\usepackage{graphicx}
\DeclareGraphicsExtensions{.eps, .pdf, .jpg, .tif}
\usepackage{bookmath}
\usepackage{mathfrak}
\usepackage{mymatrices}
\usepackage{mygroups}
\renewcommand{\onlinecite}[1]{\hspace{-1 ex} \nocite{#1}\citenum{#1}}
\def\eplus{\ve^{(+)}}
\def\eminus{\ve^{(-)}}
\def\e0{\ve^{(0)}}
\setcitestyle{numbers,super}
\begin{document}
\title{On the Excitations of a Balian-Werthamer Superconductor}
\author{J.A. Sauls} 
\affiliation{
Center for Applied Physics \& Superconducting Technologies,
Department of Physics and Astronomy, 
Northwestern University, 
Evanston, Illinois 60208}
\date{\today}
\begin{abstract}
My contribution to this collection of articles in honor of David Lee and John Reppy on their 90th birthdays is a reflection on the remarkable phenomenology of the excitation spectra of superfluid \He, in particular the B-phase which was identified by NMR and acoustic spectroscopy as the Balian-Werthamer state shown in 1963 to be the ground state of a spin-triplet, p-wave superconductor within weak-coupling BCS theory.
The superfluid phases of \He\ provide paradigms for electronic superconductors with broken space-time symmetries and non-trivial ground-state topology.    
Indeed broken spin- and orbital rotation symmetries lead to a rich spectrum of collective modes of the order parameter that can be detected using NMR, acoustic and microwave spectroscopies.
The topology of the BW state implies its low-temperature, low-energy transport properties are dominated by gapless Majorana modes confined on boundaries or interfaces.
Given the central role the BW state played I discuss the acoustic and electromagnetic signatures of the BW state, the latter being relevant if an electronic analog of superfluid \Heb\ is realized.
\end{abstract}
\maketitle

\vspace{-3mm}
\section{Introduction}
\vspace{-3mm}

In 1963 Balian and Werthamer (BW) published a paper in Physical Review based on the relatively new weak-coupling BCS theory of superconductivity with their predictions for Cooper pairing in a spin-triplet ($S=1$), orbital p-wave ($L=1$) state.~\cite{bal63} The ground state they obtained - the ``BW state'' - is \emph{isotropic } with respect to \emph{joint} rotations of the spin and orbital basis states, i.e. a state with total angular momentum $J=0$ and an isotropic excitation gap everywhere on the Fermi surface. As a result BW concluded that their ground state ``to be completely equivalent thermodynamically to the BCS state.'' and that ``The state exhibits the conventional Meissner effect, and cannot be experimentally distinguished from the BCS state by means of electromagnetic or tunneling measurements, acoustic attenuation, or nuclear magnetic resonance (NMR) relaxation times.''
A bit more than a decade later NMR and acoustic spectroscopies of the newly discovered superfluid phases of liquid \He\ led to definitive identifications of the A- and B-phases of \He\ - the A-phase as the Anderson-Morel state exhibiting broken mirror and time-reversal symmetry,\cite{and61} and the B-phase as the realization of the BW state.~\cite{leg75,and78,lee78,vollhardt90}
The theoretical framework and interpretation of both NMR frequency-shift spectroscopy and acoustic resonance spectroscopy is rooted in the theory of the collective excitations of the Cooper pair condensate.

\vspace*{-5mm}
\section{Early history of Collective Modes} 
\vspace*{-3mm}

The role of collective modes in the electromagnetic response of superconductors is a subject that originated with questions raised about gauge invariance in the context of the theory of superconductivity proposed by Bardeen, Cooper and Schrieffer (BCS).\cite{bar57}
Shortly thereafter Anderson and Nambu independently provided gauge-invariant formulations of the pairing theory and elucidated the role of the collective modes in BCS superconductors.\cite{and58a,nam60}
These modes fall into two classes: (i) \emph{massless} Nambu-Goldstone (NG) modes associated with spontaneously broken continuous symmetries, and (ii) {\it massive} Higgs modes corresponding to Bosonic excitations of the condensate.

In conventional superconductors, those in which only global $\gauge$ symmetry is spontaneously broken, the NG mode is the {\it phase} of the condensate amplitude.
This mode reflects the degeneracy of the order parameter under time-independent and spatially uniform changes in the phase. In charge neutral superconductors the phase mode is the gapless, low-energy excitation of the condensate.
The phase mode was predicted by Anderson \cite{and58} and Bogoliubov, Tolmachev and Shirkov,~\cite{bogoliubov58} and is central to spectroscopic studies of the excitation spectrum of superfluid $^3$He.~\cite{sau00a}
In particular, the Anderson-Bogoliubov (AB) mode is manifest as collisionless sound in superfluid $^3$He, and exhibits a dispersion relation, $\omega=c_{\varphi} q$, where the velocity $c_{\varphi}$, approaches the hydrodynamic sound velocity, $c_1$, of normal liquid $^3$He in the limit $T\rightarrow 0$ and $\omega\ll\Delta$.\cite{lar63}
However, for a charged superconductor the coupling to the gauge field, $\vA(\vr,t)$, and a gauge fixing condition eliminates the NG mode leaving only the dynamics of the gauge field which now obeys a Klein-Gordon equation in the long-wavelength limit, i.e. the gauge field acquires a mass given by the $M_{\text{A}}=\hbar/c\Lambda$, where $\Lambda$ is the zero-temperature London penetration length. This is the well known Anderson-Higgs mechanism for mass generation of an otherwise massless gauge field.\cite{and63,hig64}

Soon after the publication of the BCS theory several authors predicted the existence of collective modes with energies $\hbar\omega < 2 \Delta$, corresponding to excited, bound states of Cooper pairs.\cite{and58,bogoliubov58,tsu60,vak61,bar61}
Early interest in these pair excitons was stimulated by measurements of the infrared conductivity which showed a broad absorption peak onsetting well below the pair breaking edge at $\hbar\omega\approx\Delta$ in Pb and Hg films.\cite{gin60,ric60} 

However, theoretical analysis of the EM absorption, which followed soon thereafter, showed that the binding energy of these pair excitons is generally small compared with $\Delta$.\cite{tsu60} A pair exciton with energy well below the pair-breaking edge in a conventional s-wave superconductor requires a pairing interaction {\it $g_l$}, binding Cooper pairs with relative angular momentum {\it l}, that is nearly as attractive as the s-wave pairing interaction $g_0$ that binds pairs in the ground state. This was shown theoretically to be the case by Tsuneto,\cite{tsu60}, Vaks, Galitski and Larkin,\cite{vak61} and Bardasis and Schrieffer.\cite{bar61}
The excitation energies $\varepsilon_l$ for pair excitons of angular momentum $l$ are functions of $1/g_0 - 1/g_l$.\cite{bar61} Note that the exciton energies are located near 2$\Delta$ except for $g_l \simeq g_0$. The existence of two nearly degenerate pairing channels is an unlikely occurrence unless there is a weakly broken symmetry that is lifting an otherwise symmetry-protected degeneracy.~\cite{hes89} For conventional superconductors in which pairing is mediated by the electron-phonon interaction the dominant pairing channel is typically the identity representation. A near degeneracy is then unlikely except perhaps in highly anisotropic superconductors with competing Fermi surface instabilities, e.g. in the charge density or spin density channels.\cite{hes89f}
In addition, the binding of higher angular momentum pair excitons is easily destroyed by impurity scattering. The bottom line is that pair exciton modes associated with a sub-dominant pairing channel have so far not been definitively observed in any superconductor. For a critque of the early literature on collective excitations in superconductors see P. Martin in Vol. I of Park's anthology on superconductivity.\cite{parks69I}

\vspace{-3mm}
\section{Unconventional BCS pairing}
\vspace{-3mm}

In unconventional superconductors Bosonic modes with energies (masses) well separated from the unbound pair continuum edge at $2\Delta$ (hereafter ``Higgs modes'') are possible {\it without} the need for a second, nearly degenerate pairing channel. 
The reason is that, in contrast to conventional s-wave superconductors described by a complex scalar order parameter, unconventional superconductors are defined by an order parameter that breaks the maximal symmetry group of the normal state, in addition to $\gauge$ gauge symmetry.
Here I identify the order parameter with the mean-field pairing energy defined in terms of the Cooper pair amplitude, $\Delta_{\sigma\sigma'}(\vk) = g\,\langle\,a_{\vk\sigma}\,a_{-\vk\sigma'}\,\rangle$, where $a_{\vk\sigma}$ is the annihilation operator for conduction electrons with momentum $\vk$  near the Fermi surface and spin projection $\sigma\in\{\uparrow,\downarrow\}$ and $g$ is the interaction strength in the dominant attractive pairing channel. For metals with inversion symmetry the pairing interaction separates into even-parity, spin-singlet and odd-parity, spin-triplet classes in order to accomodate both parity and Fermion anti-symmetry. Thus, the order parameters for these two classes can be expressed in terms of the anti-symmetric and symmetric Pauli matrices, respectively,
\be
\Delta_{\sigma\sigma'}(\vk) = 
\left\{
\begin{array}{ll}
\Delta_{\vk}\,\left(i\sigma_y\right)_{\sigma\sigma'}
& 
,\quad\mbox{singlet} (S=0)
\,,
\cr
\vec\Delta_{\vk}\,\cdot\,\left(i\vec{\sigma}\sigma_y\right)_{\sigma\sigma'}	
& 
,\quad\mbox{triplet} (S=1)
\,.
\end{array}
\right.
\ee  
Note that $\Delta_{\vk}=\Delta_{-\vk}$ is an even-parity scalar with respect to spin rotations while the odd-parity order parameter, $\vec{\Delta}_{\vk}=-\vec{\Delta}_{-\vk}$ transforms as a vector under rotations in spin space. The dimensionality $d_{\Gamma}$ of the order parameter space is then determined by the dimensionality of the irreducible representation for the dominant pairing channel. For conventional superconductors the order parameter belongs to the identity representation; thus $d_{\Gamma}=2$ corresponding to the amplitude and phase of the complex scalar amplitude $\Delta_{\vk}$.
For unconventional superconductors the breaking of rotational symmetry implies that the order parameter, belongs to a non-trivial representation of the symmetry group of the normal state.
When $\Delta_{\vk}$ or $\vec{\Delta}_{\vk}$ belong to a higher dimensional irreducible representation of the maximal symmetry group this leads to a spectrum of {\it pairing states} belonging to this representation, and thus, bound by the same pairing interaction. This feature of the pair excitation spectrum is well studied in the case superfluid $^3$He, which is an unconventional superfluid with an order parameter that breaks rotational symmetry in both spin- and orbital space.\cite{mak74,ser74,wol78a,sau81}

Electronic analogs of the superfluid phases of \He, e.g. \upt, \ute, \sro, \cubise, $\ldots$, which exhibit multiple superconducting phases and/or belong to a higher dimensional representation of the maximal symmetry group, will host a spectrum of Bosonic excitations of the order parameter that may play an important role in the electromagnetic and acoustic response of these superconductors, at least in the clean limit $\Delta\gg \hbar/\tau$, where $\tau$ is the mean collision time for quasiparticles in the normal state.\cite{hir89,yip92,sau15,uem19}

\vspace{-5mm}
\subsection{Spin-Triplet, P-wave Pairing States}
\vspace{-3mm}

Here I consider the odd-parity/spin-triplet Balian-Werthamer (BW) state as a model for superconductivity belonging to a higher dimensional irreducible representation of the maximal symmegry group of the normal metallic state. This phase was originally investigated in the context of Knight shift measurements on elemental superconductors (Hg,V,Sn) which indicated little or no suppression of the conduction electron spin susceptibility below $T_c$.\cite{bal63}
While it was quickly realized that spin-triplet pairing was not the explanation of the temperature-independent Knight shift in these metals,\cite{abr62} the BW state has an important role in the historical development of spin-triplet pairing, and was later realized as the ground state of superfluid \He, a strongly correlated Fermi system with pairing mediated by the exchange of {\it paramagnons}, i.e. persistent ferromagnetic (FM) spin fluctuations.\cite{lay74,leg75}
Indeed \He, and thus the BW state, serves as a paradigm for superconductivity mediated by long-lived spin fluctuations, with Cooper pairs belonging to a higher dimensional representation of the symmetry group of the normal metallic state, in this case the $S=1$, $L=1$ representation of ${\sf G}=\spin\times\orbital\times\gauge\times\parity\times\time$. Thus, an electronic realization of the BW state could be a strongly correlated metal belonging to the cubic point group ($\point{O}{h}$) with FM-enhanced spin susceptibility and long-lived FM spin-fluctuations.

As noted earlier Balian and Werthamer argued that their ground state was not particularly novel in terms of its quasiparticle spectrum, thermodynamics, Meissner screening, etc. However, they did highlight one distinctive feature: ``the addition of [non-magnetic] impurities is found to reduce the critical temperature sharply. Thus, the experimental observation of the p-wave pair state is expected to be difficult ... it is suggested that a similar effect in \He\ might explain why the predicted superfluid phase has not been observed.'' This was a decade before the discovery of superfluidity in liquid \He. Fortunately, liquid \He\ is the purest form of matter in the universe!
Just as remarkable is that superfluidity, and in particular the BW state, survives the random impurity potential, albeit with a suppressed transition temperature, when liquid \He\ is infused into high-porostiy silica aerogel.\cite{thu98,hal19}

Balian and Werthamer based their conclusions primarily on the existence of an isotropic gap in the quasiparticle spectrum. They did not consider the spectrum of pair excitations, i.e. the Bosonic modes of the BW state. Nor did they anticipate that the BW ground state was a topological superconductor with topologically protected Majorana edge modes.\cite{volovik03,vol09a,miz12a,wu13}
In what follows I discuss the electrodynamcis of a BW superconductor, i.e. a charged version of superfluid \Heb. I calculate the current response to electromagnetic radiation incident on the surface of a BW superconductor, and highlight the roles and relevance of the NG and Higgs modes to the electrodynamics both above and below the pair-breaking edge of the unbound quasiparticle continuum.
I also highlight the role of the topologically protected Majorana edge states on the static EM response, specifically the contribution of Majorana modes to the temperature dependence of the London penetration depth.

Any pairing state belonging to the spin-triplet, p-wave manifold can be expressed in terms of the vector representations of $\spin\times\orbital$. In particular the spin-components of the order parameter form a $2\times 2$ matrix function of the p-wave orbital basis states, $\{\hat\vk_x,\hat\vk_y,\hat\vk_z\}$, that can be expressed as 
\be
\hspace*{-5mm}
\widehat\Delta(\vk)
\ns=\ns
\begin{pmatrix}
i\sigma_x\sigma_y
\\
i\sigma_y\sigma_y
\\
i\sigma_z\sigma_y
\end{pmatrix}^{\ns\ns\mbox{\tiny T}}
\ns\times\ns
\begin{pmatrix}
A_{xx} &
A_{xy} &
A_{xz} 
\\
A_{yx} &
A_{yy} &
A_{yz} 
\\
A_{zx} &
A_{zy} &
A_{zz} 
\end{pmatrix}
\ns\times\ns
\begin{pmatrix}
\hat\vk_x
\\
\hat\vk_y
\\
\hat\vk_z
\end{pmatrix}
\,,
\ee
where the order parameter amplitudes, $A_{\alpha i}$, define a $3\times 3$ matrix that transforms as a vector with respect to the index $\alpha\in\{x,y,z\}$ under rotations in spin space, {\sl and} separately as a vector with respect to the index $i\in\{x,y,z\}$ under space rotations. The dimensionality of the order parameter manifold is then 18 degrees of freedom that can be realized as Cooper pair excitations of the BW ground state. 
The dynamical equations describing the pair excitations are obtained from the non-equilibrium theory for the coupled equations for the distribution functions and spectral functions for quasiparticles and Cooper pairs.\cite{sau81,yip92,sau17,uem19} A key feature of this spectrum is determined by the symmetry of the BW ground state. 

\begin{center}
\begin{table}[h]
\begin{minipage}{\columnwidth}
\begin{tabular}{c|c|c|l}
		Mode	&	Symmetry	&	Energy &	Name	
\\
\hline
	$D^{(+)}_{0,M}$		&	$J=0$, $\charge=+1$	&	$2\Delta$		&	Higgs Mode	
\\
	$D^{(-)}_{0,M}$		&	$J=0$, $\charge=-1$	&	$0$			& 	NG Phase Mode	
\\
\hline
	$D^{(+)}_{1,M}$		&	$J=1$, $\charge=+1$	&	$0$			&	NG Spin-Orbit Modes
\\
	$D^{(-)}_{1,M}$		&	$J=1$, $\charge=-1$	&	$2\Delta$		&	AH Spin-Orbit Modes
\\
\hline
	$D^{(+)}_{2,M}$		&	$J=2$, $\charge=+1$	&	$\sqrt{\nicefrac{8}{5}}\Delta$	
																	&	 $2^{+}$ AH Modes
\\
	$D^{(-)}_{2,M}$		&	$J=2$, $\charge=-1$	&	$\sqrt{\nicefrac{12}{5}}\Delta$	
																	&	 $2^{-}$ AH Modes
\\
\hline
\end{tabular}
\caption{
Collective modes of the Balian-Werthamer State. The eigenmodes are labelled by 
the quantum numbers for total pair angular momentum, $J=0,1,2$, the projection 
along an axis, $m=\{-J,\ldots,+J\}$, and charge conjugation parity, 
$\charge=\{-1,+1\}$. The excitation energies for $\vq=0$ have degeneracy 
$(2J+1)$ reflecting the rotational invariance of the BW ground state.
}
\label{table-BW_modes}
\end{minipage}
\end{table}
\end{center}

\vspace{-3mm}
\subsection{The Balian-Werthamer State}
\vspace{-3mm}

The BW ground state is defined by $A^{\text{BW}}_{\alpha i} = \Delta\,\delta_{\alpha i}$. Equivalently, $\vec{\Delta}_{\vk} = \Delta\,\hat\vk$, corresponding to Cooper pairs with spin projection $\hat\vk\cdot\vS_{\text{pair}} = 0$ for any direction of orbital momentum $\hat\vk$.
Thus, the BW ground state, which breaks both rotational symmetry in spin and orbital spaces, is {\it invariant} under joint spin and orbital rotations, i.e. the BW order parameter is a condensate of $S=1$, $L=1$ Cooper pairs with total angular momentum $J=0$. Note also that the BW state breaks parity, but is time-reversal invariant. Equivalently, the residual symmetry of the BW state is ${\sf H}=\point{SO(3)}{J}\times\time$, where $\vJ=\vL+\vS$ is the generator for joint spin and orbital rotations and $\time$ is the operation of time reversal.
One manifestation of the $\point{SO(3)}{J}\times\time$ symmetry of the BW state is that the {\sl quasiparticle} excitation spectrum is rotationally invariant and degenerate with respect to spin,
\be
E_{\vk,s}=\sqrt{\xi_{\vk}^2+|\Delta|^2}\,,\quad s\in\{+\nicefrac{1}{2},-\nicefrac{1}{2}\}
\,, 
\ee
where $\xi_{\vk}=v_f(|\vk|-k_f)$, $s=\pm \nicefrac{1}{2}$ are the two helicity eigenstates, $\vec{\Delta}_{\vk}=\Delta\,\hat\vk$ and thus $\vec{\Delta}_{\vk}\cdot\vec{\Delta}_{\vk}^{*} = |\Delta|^2$, defines the excitation gap, a result highlighted by BW.\cite{bal63} Note that $\Delta=|\Delta|\,e^{i\varphi}$ is complex reflecting the broken $\gauge$ symmetry. 
The isotropy of the quasiparticle spectrum of the BW state allows us to carry out many calculations of the electromagnetic response functions analytically.\footnote{Many features of the analysis to follow are semi-quantitatively correct for {\it anisotropic} unconventional superconductors.}
There is a degenerate manifold of BW states. The general form is $\vec{\Delta}_{\vk}=\Delta\,\underline{R}[\hat\vn,\vartheta]\cdot\hat\vk$, where $\underline{R}[\hat\vn,\vartheta]$ is an orthogonal matrix which rotates the orbital states relative to the spin states by angle $\vartheta$ about the axis $\hat\vn$. I have chosen the diagonal representative with $\underline{R}=\underline{1}$ for simplicity. In nearly all that follows there is no loss in generality. An exception is the role of the nuclear dipolar energy for \Heb\ in partially lifting this degeneracy discussed in the next section.

\vspace*{-5mm}
\subsection{Collective Modes of $^3He-B$}\label{sec-Modes-3He}
\vspace*{-3mm}

Beyond the quasiparticle spectrum there is a much richer spectrum of sub-gap collective excitations of the Cooper pair condensate. 
The dimensionality of the $L=1$, $S=1$ order parameter manifold is manifest in terms of 18 collective modes. 
The fact that the BW state is invariant under joint rotations of the spin and orbital degrees of freedom implies that the Cooper pair excitations are eigenstates of $J^2,J_z$, with $J\in\{0,1,2\}$ and $J_z=M\in\{-J, \ldots, J\}$. In addition, there is a doubling of the mode spectrum associated with symmetry of the normal state under charge conjuation, i.e. the transformation of conduction electrons into holes and vice versa.\cite{ser83a,sau84a,fis85}
Table \ref{table-BW_modes} summarizes the symmetries and excitation energies of the collective mode spectrum of the BW state in the long-wavelength limit.
There are 4 gapless Nambu-Goldstone modes associated with the continuous degeneracy space, 
\be
R=\frac{
\gauge
\times
\spin\times\orbital
}
{\point{SO(3)}{L+S}}
=
\gauge\times\point{SO(3)}{L-S}
\,.
\ee
The NG modes include the phase mode associated with the $\gauge$ degeneracy, the $J=0$, $\charge=-1$ mode in Table~\ref{table-BW_modes}.
In addition there are 3 NG modes associated with the $\point{SO(3)}{L-S}$ degeneracy resulting from broken \emph{relative} spin-orbit rotation symmetry, total angular momentum $J=1$ and charge conjugation parity $\charge=+1$, also listed in Table~\ref{table-BW_modes}. These are spin-orbit NG modes.

In the neutral superfluid the phase mode is realized as collisionless sound, and plays a central role in acoustic spectrosopy of \Heb\ as discussed in Sec.~\ref{sec-Higgs-3He}. In the case of a charged condensate the phase mode can be removed by a gauge fixing condition, and the resulting dynamics is that of the gauge field which acquires a mass as discussed in Sec.~\ref{sec-Anderson-Higgs_Mechanism}.

The $J^{\charge}=1^{+}$ NG modes are spin-orbit waves with excitation energies, $\hbar\Omega_{1,m}=c_{1,m}|\vq|$, and velocities, $c_{1,0}=\nicefrac{1}{5}v_f$ and $c_{1,\pm 1}=\nicefrac{2}{5}v_f$ in the weak-coupling limit.\cite{popov87}
The weak breaking of \emph{relative} spin-orbit rotation symmetry by the nuclear dipolar interaction \emph{above $T_c$} partially lifts the degeneracy of the $J^{\charge} = 1^{+}$ NG modes, endowing the $M=0$ mode with a small gap (mass) determined by the nuclear dipole energy, $M_{\text{LH}}=\hbar\Omega_{\text{B}}\ll\Delta$. This mode was first obtained by Leggett in his development of the spin dynamics of \He\ including both spontaneously broken spin-orbit rotation symmetry and partial lifting of the 3-fold degeneracy by the weak nuclear dipolar energy.~\cite{leg74} 
The result is a pseudo-NG triplet with a gapped $J=1^{+},M=0$ mode and two gapless modes with $M=\pm 1$. The $M=0$ mode is observable as a \emph{longitudinal} NMR resonance at $\omega=\Omega_{\text{B}}$, while in linear response transverse NMR exhibits no shift from the Larmor frequency. These results, and the spin dynamics of Leggett's equations under large amplitude excitation, played a central role in identifying \Heb\ as the BW state.\cite{leg75}
More recently, the $J^{\charge}=1^{+}$ multiplet was argued to provide a novel example of mass generation in quantum field theory corresponding to the ``Light Higgs'' extension of the standard model in particle physics.\cite{zav16}
A direct detection of the $J=1^{+},M=0$ Light Higgs Boson in \Heb\ was achieved by measuring the decay of optical magnons created by magnetic pumping (a magnon BEC). A sharp threshold for decay of optical magnons to a pair of Light Higgs modes was observed by tuning the mass of the optical magnons on resonance, i.e. $M_{\text{opt}} = \hbar\gamma B\ge 2M_{\text{LH}}=2\hbar\Omega_{\text{B}}$.\cite{zav16}

The remaining 14 modes are gapped excitations of the condensate (Higgs modes) that couple to charge currents, spin currents or energy density.

\begin{figure}[t]
\includegraphics[width=0.95\linewidth,keepaspectratio]{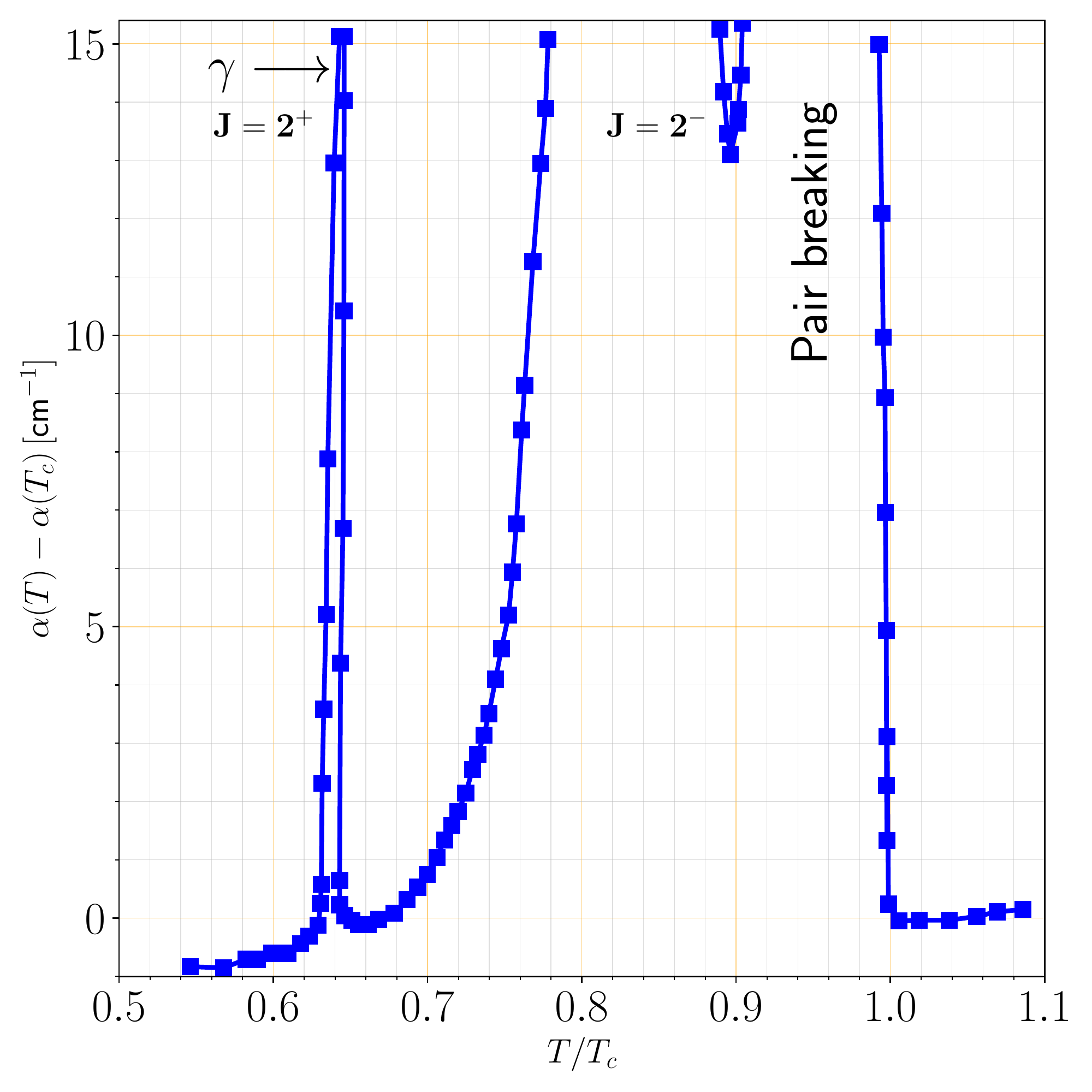}
\caption{\small
Attenuation of zero sound in \Heb\ as a function of temperature at frequency $\omega/2\pi= 60\,\mbox{MHz}$ and a pressure of $p=5.3\,\mbox{bar}$. The peak labelled $\gamma$ is identified as absorption of zero-sound phonons by resonant excitation of the $J=2^{+}$ Higgs mode. Much stronger absorption results from resonant excitation of the $J=2^{-}$ mode and pair breaking for $\hbar\omega\ge 2\Delta(T)$. 
Figure created from data in Giannetta et al.~\cite{gia80}
}
\label{fig-absorption_resonance_Cornell}
\end{figure}

\vspace*{-5mm}
\subsection{Acoustic Spectroscopy of $^3He-B$}\label{sec-Higgs-3He}
\vspace*{-3mm}

The collective modes of the BW state were studied theoretically soon after the discovery of the superfluid phases of \He,~\cite{wol74,ser74,mak74} The modes were discovered theoretically much earlier by Vdovin,~\cite{vdo63} however, see Sec.~\ref{sec-Vdovin} on my critique of Vdovin's paper.
More extensive studies of the collective modes followed,~\cite{koc81,sau81,sau82c,sch83} driven by the discovery of a sharp resonance in the absorption spectrum of zero sound below the pair-breaking continuum at $\hbar\omega_{\gamma}\simeq 1.1\Delta(T)$ by the low-temperature (LT) group at Cornell headed by David Lee, and independently by the Northwestern LT group headed by William Halperin.~\cite{gia80,mas80}
The narrow resonance ($\gamma$) in the attenuation spectrum reported by the Cornell group is reproduced from data of Ref.~\onlinecite{gia80} in Fig.~\ref{fig-absorption_resonance_Cornell}. The spectrum also shows much larger contributions to the attenuation from pair breaking at temperatures satisfying $\hbar\omega\ge 2\Delta(T)$, as well as high attenuation at frequencies near the $J=2^{-}$ mode.   
This discovery led to a flurry of investigations aimed at identifying the origin of the $\gamma$ peak, and the mechanism providing the coupling of the mode to zero sound phonons. 
Based on theoretical results published prior to 1980 it was expected that the $J=2^{-}$ Higgs mode could be excited by zero sound ($J=0^{-}$) leading to resonant absorption of zero-sound phonons, but the $J=2^{+}$ Higgs modes were de-coupled from sound, but would couple to excitations of the nuclear spin or spin current.~\cite{wol74,ser74,mak74} 

Several proposals were put forward for the absorption resonance. Sauls and Serene showed that a sub-dominant f-wave pairing interaction would lead to a pair exciton mode with $S=1$, $L=3$, $J=4$ and $\charge=-1$ that satisfies the selection rules for coupling to zero sound phonons, but with significantly reduced coupling strength. However, identifying the $J=4^{-}$ mode with the $\gamma$ resonance would require a sufficiently strong attractive f-wave pairing interaction for the pair exciton to be bound with energy $\hbar\omega_{4^{-}}\approx 1.07\Delta(T)$.~\cite{sau81}
At about the same time Koch and W\"olfle proposed a mechanism to lift the selection rule preventing the coupling of the $J=2^{+}$ mode to zero sound. Their theory introduced weak particle-hole asymmetry of the {\it normal-state} density of states, i.e. $\eta=T_c N^{'}(0)/N(0)\approx 10^{-3}$. This asymmetry allows for a weak coupling ($\cO(\eta)$) of the $J=2^{+}$ Higgs mode to the density and current density of zero sound.~\cite{koc81} 
The experimental identification of the $\gamma$ resonance in favor of the $J=2^{+}$ Higgs modes was the observation of a five-fold splitting of the absorption resonance in a magnetic field by Avenel and Varoquaux, corresponding to the lifting of the $(2J+1)$ degeneracy of $J=2^{+}$ nuclear Zeeman multiplet.~\cite{ave80}

\begin{figure}[t]
\includegraphics[width=0.95\linewidth,keepaspectratio]{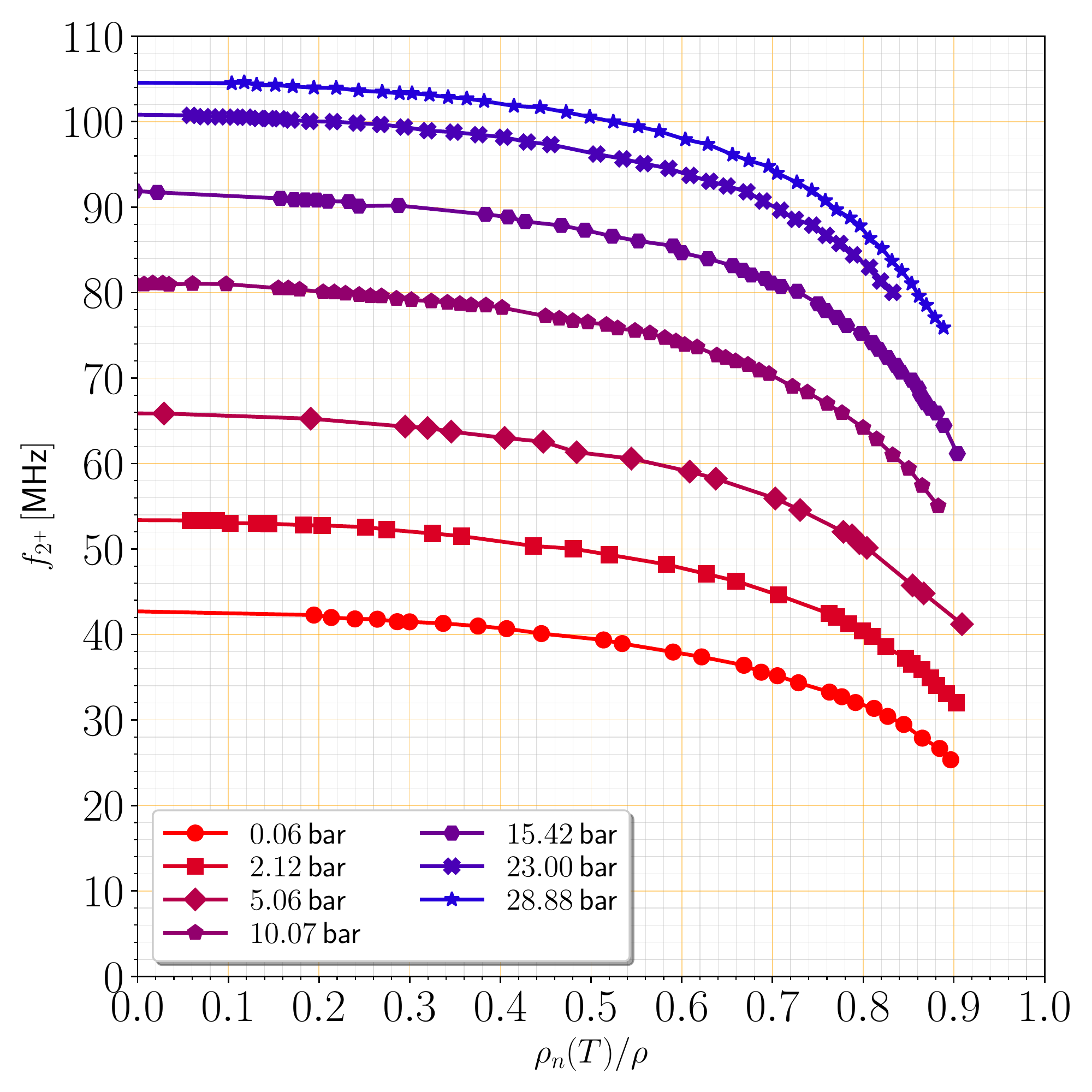}
\caption{\small
The $J=2^{+}$ Higgs mass, $f_{2^+} = M_{2^+}\,c^2/h$, expressed in $\mbox{MHz}$ as a function of $\rho_n(T)/\rho$ (temperature) and pressure. Figure produced from data reported in Fraenkel et al.~\cite{fra89}
}
\label{fig-J=2+_frequency_vs_pressure}
\end{figure}

Nevertheless, a complication was that the weak-coupling prediction for the Higgs mass, $M_{2^{+}}^{\text{wc}}=\sqrt{\frac{8}{5}}\Delta\simeq 1.265\Delta$ was substantially larger than the observed mass of $M_{\gamma}\approx 1.07\,\Delta$. The discrepancy between the weak-coupling prediction for the mass of the $J=2^{+}$ mode and the observed resonance was resolved by the additional binding from the f-wave pairing channel that, in addition to supporting sub-gap $J=4^{\pm}$ modes, also allowed for an $L=3$, $S=1$, $J=2^{+}$ amplitude that mixes with the $L=1$, $S=1$, $J=2^{+}$ amplitude leading to an f-wave pairing correction to the $J=2^{+}$ mass.~\cite{sau81} 
Experimental measurements of the mass shift of the $J=2^{+}$ mode were made by several groups. The results for the mass of the $J=2^{+}$ mode reported by the Cornell group headed by John Reppy are shown in Fig.~\ref{fig-J=2+_frequency_vs_pressure}.~\cite{fra89} These measurements were analyzed and imply substantial corrections to the weak-coupling result from f-wave pairing interactions based on the theory of Ref.~\onlinecite{sau81}.
A comprehensive review of the theory of the mass spectrum of the NG and Higgs modes of \He\ is published in Ref.~\onlinecite{sau17}.

The discovery by Giannetta et al.~\cite{gia80} and Mast et al.~\cite{mas80} set off a multi-decade research program to understand the Bosonic spectrum of superfluid \He\ that led to many remarkable discoveries, including the predictions~\cite{sau82c,sch83,fis86,fis88a} and discoveries~\cite{ave80,shi83} of the nuclear Zeeman and Paschen-Bach effects on the mode spectrum, the observation soliton propagation of zero sound~\cite{pol81a} mediated by coherent transitions between the $J=0^{+}$ ground state and the $J=2^{+}$ Higgs level,~\cite{sau81d} the prediction~\cite{mck89a,mck90,sau91,mck92} and discovery of three-wave mixing and two-phonon absorption of zero sound by the $J=2^{+}$ modes,~\cite{tor92,man94}, the prediction of transverse zero sound mediated by the $J=2^{-},M_J=\pm 1$ modes, as well as acoustic Faraday rotation of the mass current polarization as a direct signature of spontaneously broken relative spin-orbit rotation symmetry.~\cite{moo93,sau00b} The theory was confirmed by the discovery of propagating transverse sound at frequencies above the threshold set by the mass of the $J=2^{-}$ modes, and the observation of Faraday rotation of the mass current polarization for magnetic fields along the direction of propagation of transverse sound.~\cite{lee99,col13} Transverse sound and impedance spectroscopy led to the discovery of a new collective mode with excitation energy just below the pair-breaking threshold, $\hbar\omega\lesssim 2\Delta$,~\cite{dav08} that is consistent with the predicted $J=4^{-},M=\pm 1$ modes.\cite{sau81}. These are just a few of the discoveries that grew from the discovery of the $J=2^{+}$ mode using ultrasound spectroscopy. For an indepth look see the reviews by Halperin and Varaquax~\cite{hal90}, McKenzie and Sauls,~\cite{mck90} and Sauls.~\cite{sau00a}

\subsection{Collective Modes of Superconductors}
\vspace{-3mm}

Unconventional superconductors, e.g. \upt, are often type II superconductors with Meissner penetration lengths that are large compared to the coherence length, $\Lambda\gg\xi$.\cite{sig95b} Thus, the electromagnetic field penetrates relatively deep into the superconductor at a vacuum/superconducting interface, and probes the {\it bulk} order parameter by exciting currents far from the interface where the order parameter is often distorted from its bulk form on the scale of the corherence length. Thus, in what follows I negelect the surface deformation of the order parameter in calculating the current response. For weak EM fields the induced current is linear in the field $\vA(\vq,\omega)$,
\be\label{eq-linear_current_response}
\vJ(\vq,\omega)=-\mathbb{K}(\vq,\omega)\cdot\vA(\vq,\omega)
\,.
\ee
where $\mathbb{K}(\vq,\omega)$ is a second-rank tensor representing the current response to the EM field $\vA(\vq,\omega)$. This response function includes the unbound quasiparticle contribution to the response to the EM field as well as the response of condensate, including any contributions from order parameter collective modes to the charge current.
To calculate $\mathbb{K}(\vq,\omega)$ I solve the quasiclassical transport equations for the Keldysh propagator in Nambu space, $\whmfGk(\vk,\varepsilon;\vq, \omega)$, for quasiparticles and pairs defined by momentum $\vk$ on the Fermi surface with excitation energy $\varepsilon$ in the presence of an EM field $\vA(\vq,\omega)$. The latter is determined self-consistently from Maxwell's equation with the induced current source $\vJ(\vq,\omega)$.
The current response is then given in terms of the of the $\tauz$-component the Keldysh propagator,\cite{hir89,yip92,sau17,uem19}
\be\label{eq-Current-response}
\vJ(\vq,\omega)=N_f\int\,\frac{d\Omega_{\vk}}{4\pi}\left(-e\vv_{\vk}\right)
\int \frac{d\varepsilon}{2\pi i}\,\mfG(\vk,\varepsilon;\vq,\omega)
\,,
\ee 
where $\mfG(\vk,\varepsilon;\vq,\omega)$ is the solution of the linearized quasiclassical Keldysh transport equation. Note that $-e$ is the electron charge. In the clean limit the energy integrated Keldysh propagator is given by
\begin{widetext}
\be\label{eq-Gk}
\hspace*{-3mm}
\int\ns\frac{d\varepsilon}{2\pi i}\,\mfG(\vk,\varepsilon;\vq,\omega)
\ns=\ns 
\frac{2e}{c}\lambda(\eta,\omega)\left(\vv_{\vk}\cdot\vA\right) 
\ns-\ns 2ie \left(\frac{\omega}{\omega^{2} - \eta^{2}}\right)
\,\left(1-\lambda(\eta,\omega)\right)
  \left(\vv_{\vk}\cdot\vE\right) 
\ns+\ns
\lambda(\eta,\omega)\left(\frac{\eta}{|\Delta|^{2}}\right)
\left[\vec{\Delta}_{\vk}\cdot\vec{d}^{(-)}_{\vk}\right]
\,,
\ee
\end{widetext}
where $\eta=\vv_{\vk}\cdot\vq$ and
\be\label{eq-tsuneto_function}
\hspace*{-2mm}
\lambda(\eta,\omega)
\ns=\ns
|\Delta|^2\ns\int\limits_{-\infty }^{+\infty}
\ns
\frac{d\varepsilon}{2 \pi i} 
\left[\frac{2\varepsilon\omega\,D_{\beta}(\varepsilon;\omega) 
    + \eta^{2}\,S_{\beta}(\varepsilon;\omega)}{D\left(\varepsilon;\eta,\omega\right)}
\right],
\ee
with $D_{\beta} = \beta_{+}-\beta_{-}$, $S_{\beta} = \beta_{+}+\beta_{-}$, 
$\beta_{\pm}= \beta(\varepsilon\pm\omega/2)$, and
\be
\beta(\varepsilon)
\equiv
2\pi i\,\frac{\sgn(\varepsilon)}{\sqrt{\varepsilon^2 - \Delta^2}}\,
        \tanh\left(\frac{\varepsilon}{2T}\right)\,
        \Theta\left(\varepsilon^{2} - \Delta^{2}\right)
\,.
\ee
The denominator is defined by
\be
D(\varepsilon;\eta,\omega) = (4\varepsilon^2-\eta^2)(\omega^2-\eta^2) + 4|\Delta|^2\eta^2
\,.
\ee
Note that I have fixed the global phase such that $\vec{\Delta}_{\vk} = \vec{\Delta}_{\vk}^{*}$. This conveniently fixes the charge conjugation parity of the BW ground state to be $\charge=+1$; hereafter I identify the gap with $|\Delta|=\Delta$. The details of this calculation are given in Ref.\onlinecite{yip92,sau17}.

The first term on the right-hand side of Eq.~\ref{eq-Gk} is the condensate response to the vector potential, $\vA$, and the second term is the quasiparticle response to the electric field, $\vE$. The former is the non-dissipative a.c. supercurrent for frequencies $\hbar\omega < 2\Delta$, while the latter encodes the dissipative response of the non-equilibrium quasiparticle excitations at finite temperature, frequency and wavelength. 
Note that for $\omega=0$ and $q\rightarrow 0$ the Tsuneto function, $\lambda(\eta,\omega)$, reduces to the equilibrium condensate stiffness, or ``superfluid fraction'', while the long-wavelength Tsuneto function, $\lambda(\omega)\equiv\lambda(0,\omega)$, is the nonequilibrium condensate stiffness for $\hbar\omega<2\Delta$.
At high frequencies, $\hbar\omega \ge 2\Delta$, the Tsuneto function acquires an imaginary part representing the spectral density of unbound quasiparticles created by dissociation of Cooper pairs by absorption of microwave photons of energy $\hbar\omega$.
The last term in Eq.~\ref{eq-Gk}, proportional to $\vec{\Delta}_{\vk}\cdot\vec{d}_{\vk}^{(-)}$, represents the non-equilibrium contribution to the Keldysh propagator from collective excitations of the condensate with charge conjugation parity $\charge=-1$ under the transformation: particle $\leftrightarrow$ hole. N.B. $\vec{d}_{\vk}^{(\pm)}=\vec{d}_{\vk}(\vq,\omega)\pm\vec{d}_{\vk}(-\vq,-\omega)^{*}$.\footnote{Particle-hole symmetry implies a selection rule: $\charge=+1$ modes do not couple to $\vv_{\vk}\cdot\vA$.\cite{ser83a,sau84a}}

\vspace{-5mm}
\subsection{Nambu-Goldstone and Higgs Modes}
\vspace{-3mm}

For superconductors described by a complex scalar order parameter the space-time dynamics of the Cooper pairs separates into excitations of the phase and amplitude of the mean field order parameter, which for small deviations from equilbrium take the form,
$\Delta(\vr,t) = |\Delta|\left(1 + h(\vr,t) + i\varphi(\vr,t)\right)$,
where $h(\vr,t)$ is a real scalar field representing the amplitude fluctuations, and $\varphi(\vr,t)$ is the corresponding phase fluctuations. 
Similarly, if the dynamics of the BW state was restricted to amplitude and phase fluctuations of the p-wave, spin-triplet ground state with $J=0$, then the dynamics of the order parameter would be described by two scalar fields, 
$\vec{d}_{\vk}^{(+)}(\vr,t) = 2\vec{\Delta}_{\vk}\,h(\vr,t)$ and 
$\vec{d}_{\vk}^{(-)}(\vr,t) = 2i\vec{\Delta}_{\vk}\,\varphi(\vr,t)$.
And just as was originally found for conventional superconductors, in the absence of the coupling to the electromagnetic gauge field, the amplitude and phase are eigenmodes of the condensate obeying the dynamical equations,
\ber
&&\left(\partial_t{^2} - c_{\varphi}^2\,\nabla^2\right)\varphi(\vr,t) = 0
\,,
\\
&&\left(\partial_t{^2} - c_{h}^2\,\nabla^2 + M_h^2\right)\,h(\vr,t) = 0
\,.
\eer
The phase mode is the massless Nambu-Goldstone mode associated with the broken $\gauge$ symmetry, and obeys a wave equation with phase velocity $c_{\varphi}=v_f/\sqrt{3}$.
The amplitude mode obeys a Klein-Gordon equation corresonding to Cooper pair excitations of mass $M_h=2|\Delta|$ and velocity $c_{h}=v_f/\sqrt{3}$. This is the Higgs excitation, which has the same quantum numbers ($S=1$, $L=1$, $J=0$, $\charge=+1$) as ground state Cooper pairs, and thus is decoupled from the EM field via single photon processes.

\vspace{-5mm}
\subsection{Gauge Invariance \& the Anderson-Higgs Mechanism}
\label{sec-Anderson-Higgs_Mechanism}
\vspace{-3mm}

For a charged superconductor the NG phase mode disappears; it can be absorbed into the gauge field, $\vA(\vr,t)$. Consider a local gauge transformation defined by the scalar field, $\chi(\vr,t)$; the potentials transform as 
\ber
\vA \rightarrow \vA'
\ns&\ns=\ns&
\vA + \grad\chi(\vr,t) \xrightarrow[]{\mbox{FT}} \vA(\vq,\omega) + i\vq\chi(\vq,\omega)
\label{eq-gauge_A}
\,,
\\
\Phi \rightarrow \Phi'
\ns&\ns=\ns&
\Phi -\frac{1}{c}\partial_t\chi(\vr,t) 
\xrightarrow[]{\mbox{FT}} 
\Phi(\vq,\omega) + i\frac{\omega}{c}\chi(\vq,\omega)
\,,
\hspace{5mm}
\label{eq-gauge_Phi}
\eer
with the right-hand side of Eqs.~\ref{eq-gauge_A}-\ref{eq-gauge_Phi} the corresponding Fourier transforms. Local gauge invariance is ensured by a corresponding change of phase of the condensate amplitude. Thus, the space-time mean-field order parameter transforms as, 
\be
\vec\Delta_{\vk}(\vr,t)\rightarrow\vec\Delta_{\vk}^{'}(\vr,t)
=
\vec\Delta_{\vk}(\vr,t)\,e^{-i\nicefrac{2e}{c}\chi(\vr,t)}
\,.
\ee
Indeed local gauge invariance is encoded in Eq.~\ref{eq-Gk} for the equal-time linear response for the Keldysh function by considering an infinitesimal gauge transformation with 
\be
\hspace*{-2mm}
\vec{d}_{\vk}^{(-)}\ns(\vq,\omega)
\ns\rightarrow\ns
\vec{d}_{\vk}^{(-)\,'}\ns(\vq,\omega)
\ns=\ns
\vec{d}^{(-)}_{\vk}\ns(\vq,\omega) - i\nicefrac{2e}{c}\chi(\vq,\omega)\vec{\Delta}_{\vk}
\,.
\ee

Thus, if we consider the NG mode of the BW ground state expressed in terms of the condensate phase, $\vec{d}_{\vk}^{(-)} = 2i\vec{\Delta}_{\vk}\,\varphi(\vq,\omega)$, then given the potentials, $\mathcal{A}_{\mu}=(\vA,-\nicefrac{1}{c}\Phi)$ with $\mu=\{1,2,3,4\}$, we {\it fix the gauge} of $\mathcal{A}_{\mu}$ to cancel the phase field $\nicefrac{e}{c}\chi(\vq,\omega)=\varphi(\vq,\omega)$, which then removes the massless NG mode from the current response in Eq.~\ref{eq-Gk}. 
In a conventional superconductor the remaining dynamics is the amplitude (Higgs) mode, which is decoupled from the EM field, and that of the gauge field which obeys Maxwell's equation with only the first two terms in Eq.~\ref{eq-Gk} contributing to the current in Eq.~\ref{eq-Current-response}. 
In the long-wavelength limit, $q v_f \ll \omega$, the transverse current reduces to 
\be\label{eq-current_q=0}
\vJ = -\frac{2}{3}N_f \frac{e^2 v_f^2}{c}\,\vA
\,.
\ee
The transverse components of the vector potential ($\dive{\vA}=0$) then satisfy the Klein-Gordon equation,
\be\label{eq-AH-mode}
\left(\partial_t^2 - c^2\nabla^2 + \omega_p^2\right)\,\vA = 0
\,,
\ee
where $\omega_p$ is the plasma frequency,
\be
\omega_p^2 = \frac{8\pi}{3}\,N_f\,e^2\,v_f^2 = \frac{4\pi\,n\,e^2}{m^*}
\,,
\ee
with carrier density, $n$, and effective mass $m^*$ of the normal-state conduction electrons. {\sl Prima facie} Eq.~\ref{eq-AH-mode} implies propagating transverse EM waves for frequencies $\omega > \omega_p$.
uantization of the EM field then implies the existence of vector Bosons with energy, 
\be
E_{\vp}=\sqrt{p^2 c^2+M_{\text{A}}^2 c^4}
\,,
\ee
and momentum, $\vp=\hbar\vq$, i.e. {\it photons acquire a mass} related to the zero-temperature London penetration length, $\Lambda=c/\omega_p$,
\be
M_{\text{A}} = \hbar/c\Lambda
\,.
\ee
This is the \emph{Anderson-Higgs mechanism} for mass generation of an otherwise massless NG Boson.\cite{and63,hig64} This is a remarkable feature of global $\gauge$ symmetry breaking and local gauge invariance. The AB mode of a neutral Cooper pair condensate plays an essential role as a propagating acoustic phonon in the collisionless limit, while the gauge fixing condition for a charged condensate, which absorbs the phase into the gauge field, eliminates the massless NG mode leaving behind a gauge Boson with mass $M_{\text{A}}=\hbar\omega_p/c^2$.

\vspace{-5mm}
\subsection{Meissner Effect \& the Gauge Boson Mass}
\vspace{-3mm}

The KG equation for the gauge field describes long-wavelength, massive gauge Bosons for excition energies, $E$, just above the plasma energy, i.e. $\hbar c q =\sqrt{E^2 -\hbar^2\omega_p^2}$. The plasma energy for nearly all superconductors is a high-energy scale compared to any energy associated with superconductivity. Thus, the propagation of massive gauge Bosons at energies above the plasma energy is of little relevance to the electrodynamics of most superconductors.

However, we can infer the {\it existence of the massive gauge Boson} by considering the static limit, $\omega=0$. There are no propagating gauge Bosons; instead the gauge field has only localized solutions corresponding to static confined magnetic fields on the scale of the London penetration depth at superconducting-vacuum interfaces.
In this limit the current and gauge field are obtained from Eqs.~\ref{eq-Current-response}-\ref{eq-Gk} with the NG mode removed by the gauge fixing condition, 

\be
\vJ(\vq) = N_f\int\frac{d\Omega_{\vk}}{4\pi}(-e\vv_{\vk})\frac{2e}{c}\,
\lambda(\vq\cdot\vv_{\vk})\,\vv_{\vk}\cdot\vA(\vq)
\,,
\label{eq-current_vs_T_omega=0}
\ee
where 
\be\label{eq-Tsuneto_omega=0}
\hspace*{-3mm}
\lambda(\eta)\ns=\ns|\Delta|^2\ns\fint_{-\infty}^{+\infty}\ns\ns\ns\ns d\xi
                 \frac{\tanh(\sqrt{\xi^2+|\Delta|^2}/2T)}{2\sqrt{\xi^2+|\Delta|^2}}
                 \ns\times\ns\frac{1}{\nicefrac{1}{4}\eta^2 - \xi^2}
\,,
\ee

\noindent is the static limit of the condensate response.\footnote{I omitted contributions from off-resonant collective modes with $J\ne0$. This is justified in the London limit $v_fq\ll\Delta$. Note that $\fint$ is principal part integration in the neighborhood of the integrable singularities at $\pm\nicefrac{1}{2}\eta$.}
The response function is readily evaluated using the Matsubara representation for 
$\tanh(\varepsilon/2T)/2\varepsilon = T\sum_{\varepsilon_n}(\varepsilon_n^2+\varepsilon^2)^{-1}$
where $\varepsilon_n=(2n+1)\pi T$ are the Fermion Matsubara frequencies with $n\in\mathbb{Z}$.
Equation~\ref{eq-Tsuneto_omega=0} can now be transformed (see Appendix) to
\be\label{eq-Tsuneto_omega=0-regular}
\hspace*{-3mm}
\lambda(\eta)\ns=
\pi T\sum_{\varepsilon_n}\frac{|\Delta|^2}{\sqrt{\varepsilon_n^2+|\Delta|^2}} 
\times\frac{1}{\varepsilon_n^2+|\Delta|^2 + \nicefrac{1}{4}\eta^2} 
\,.
\ee
In general the current response is a non-local function of the vector potential. However, if the confinment length is long compared to the coherence length, $\xi_0=\hbar v_f/2\pi T_c$, then the relevant wavevectors satisfy $q v_f \ll 2\pi T_c, |\Delta|$, in which case we can evaluate Eq.~\ref{eq-Tsuneto_omega=0-regular} in the limit $\eta\rightarrow 0$.
The result is the Yosida function for the temperature-dependent superfluid fraction, $n_s/n$. The resulting current response is now a local function of $\vA$, $\frac{4\pi}{c}\vJ(\vr) = - \frac{1}{\Lambda_{\text{L}}^2}\,\vA(\vr)$, and the gauge field then satisfies London's equation,
\be
\hspace*{-3mm}
\left(-\nabla^2 + \frac{1}{\Lambda_{\text{L}}^2}\right)\,\vA = 0
\,,\mbox{with}\,\,
\Lambda_{\text{L}}=\frac{1}{\sqrt{\lambda(\eta\rightarrow 0)}}
\frac{M_{\text{A}}c}{\hbar}
\,.
\ee
The confinment length, $\Lambda_{\text{L}}$, is determined by the the mass of the gauge Boson and the condensate response function; $\Lambda_{\text{L}}$ is the {\it temperature-dependent} London penetration depth, 
\be
\Lambda_{\text{L}}\equiv\sqrt{\frac{m^* c^2}{4\pi n_s e^2}}
\,,
\ee
where $n_s \equiv n\,\lim_{\eta \rightarrow 0}\lambda(\eta)$ is the {\it superfluid density} defined by the static, long-wavelength limit of the condensate response,
\be\label{eq-superfluid_fraction}
\hspace*{-3mm}
\frac{n_s}{n}
\ns=\ns 
\pi T\sum_{\varepsilon_n}
\frac{|\Delta|^2}
     {\left[\varepsilon_n^2+|\Delta|^2\right]^{\nicefrac{3}{2}}}
\ns=\ns 
\Bigg\{
\ns\begin{array}{ll}
1 - \sqrt{\frac{2\pi |\Delta|}{T}}\,e^{-|\Delta|/T}
& 
\ns,\,T\rightarrow 0, 
\cr 
\displaystyle{\nicefrac{7\zeta(3)}{4\pi^2}\frac{|\Delta|^2}{T_c^2}}
& 
\ns,\,T\rightarrow T_c^{-}.
\end{array}
\ee
Note that $|\Delta|$ is the weak-coupling gap with $|\Delta|=1.76 T_c$ for $T=0$, and for $T\rightarrow T_c^-$, $|\Delta|^2\approx\nicefrac{\pi^2 T_c^2}{7\zeta(3)/8}\,(1-T/T_c)$.

\vspace*{-5mm}
\subsection{Meissner Effect and Topology of the BW state}
\vspace*{-3mm}

Balian and Werthamer's prediciton that the quasiparticle spectrum is fully gapped over the Fermi surface was the basis for their conclusion that the BW ``state exhibits the conventional Meissner effect, and cannot be distinguished from the BCS state''. Equation~\eqref{eq-superfluid_fraction} highlights the weak exponential reduction of the superfluid fraction, or increase in the London penetration depth, $\delta\Lambda_{L}(T)/\Lambda\approx\sqrt{\frac{\pi|\Delta|}{2T}}\,e^{-|\Delta|/T}$, for $T\ll |\Delta|$.
However, it was not known at that time that the BW ground state is a 3D time-reversal invariant topological superfluid~\cite{volovik03,vol09a} belonging to class DIII with winding number $N_{\text{3D}}=2$, protected by time-reversal and charge-conjugation symmetry, $\Gamma=\time\times\charge$.~\cite{sch08}

At a vacuum-superconducting interface the bulk-boundary correspondence~\cite{hat93a} implies there is a spectrum of gapless Majorana modes with $\varepsilon_{\pm}(\vk)=\pm c\,|\vk_{||}|$ where $c=|\Delta|/p_f$ and $\vk_{||}$ is the momentum parallel to theinterface. The dispersion relation forms a pair Majorana cones above and below the Fermi level with the zero energy state protected by the bulk topology of the BW state.
Furthermore, the topology is preserved for the special class of current-carrying states with condensate momentum $\vp_s=\nicefrac{\hbar}{2}\left(\grad\varphi-\nicefrac{2e}{c}\vA\right)$ paralell to the vacuum-superconducting interface. Such states break $\time$ symmetry, as well as rotational symmetry about the axis normal to the interface. However, the product $\time\times\point{U}{z}(\pi)$ is a symmetry of the current-carrying BW state. N.B. $\point{U}{z}(\pi)$ is a $180^{\sf o}$ rotation about the normal to the interface. As a result the non-trivial topology of the BW state remains protected by the product of discrete symmetries, $\Gamma=\point{U}{z}(\pi)\times\time\times\charge$.~\cite{wu13}

However, the condensate flow field generates a Doppler shift of the spectrum of Majorana modes: $\varepsilon_{\pm}(\vk)=\pm c\,|\vk_{||}| - \vp_s\cdot\vv_{\vk}$, which for in-plane condensate flow are positive and negative energy Majorana cones with {\it anisotropic} velocities $c_{\pm}(\phi_{\vk})=|\Delta|/p_f\pm (|\vp_s|/m^*)\,\cos\phi_{\vk}$, where $\phi_{\vk}$ is the azimuthal direction of $\vk_{||}$ relative to $\vp_s$.
Since the condensate flow does not shift states across the Fermi energy the ground-state current is unaffected by the Majorana spectrum. However, the in-plane anisotropy in the Majorana spectrum leads to a power-law correction to the current at temperatures $T\ll |\Delta|$.
In particular for superfluid mass flow of \Heb\ confined in a channel of width $D\gg\xi_{\Delta}=\hbar v_f/\pi |\Delta|$ the leading order correction to the superfluid fraction is
\be
\frac{n_s}{n}\approx 1 - \frac{27\pi\zeta(3)}{2}\,\frac{\xi_{\Delta}}{D}\,\frac{m^*}{m}\,\left(\frac{T}{|\Delta|}\right)^3
\,.
\ee
Thus, thermal excitation of the Majorana modes leads to reduction of the superfluid fraction that scales as $T^3$, and is characteristic of the gapless linearly dispersing Majorana modes confined to the 2D interface.~\cite{wu13}

For a BW superconductor the Meissner screening current at the vacuum-superconducting interface, and thus the London penetration depth, acquires a $T^3$ correction to the zero-temperature London length, $\Lambda$, given by
\be
\frac{\delta\Lambda_{L}}{\Lambda} \approx -\frac{1}{2}\frac{\delta n_s}{n}\approx
\frac{27\pi\zeta(3)}{8}\,\frac{\xi_{\Delta}}{\Lambda}\,\frac{m^*}{m}\,\left(\frac{T}{|\Delta|}\right)^3
\,.
\ee

Note that Meissner screening confines the condensate flow to an effective thickness $D/2\approx\Lambda$. The $T^3$ correction to the London penetration depth for the class of DIII topological superconductors was also obtained by the authors of Ref.~\onlinecite{wu20}. Observation of the $T^3$ correction to the London penetration depth in a fully gapped superconductor would provide strong evidence for gapless Majorana modes on the boundary of a topological superconductor. 

\vspace*{-5mm}
\section{Electrodynamics of the BW state}
\vspace*{-3mm}

For conventional single-band, s-wave, spin-singlet superconductors, with no sub-dominant pairing channels, the Anderson-Higgs mechanism renders the NG mode irrelevant to EM fields with photon energies of order $\hbar\omega\lesssim 2\Delta$, leaving only the low-frequency Meissner response and the $J=0^+$ Higgs mode which exhausts the collective mode response of the condensate. Furthermore, the spin and parity of the Higgs mode renders it inaccessible via single photon absorption. I discuss in Sec.~\ref{sec-J=0+_Higgs} the $J=0^+$ Higgs mode which is accessible via two-photon or two-phonon processes. However, first I discuss the full Bosonic mode spectrum of the BW state. 
The results discussed in this section for the coupling to the Bosonic excitations to the EM field for electronic analogs of superfluid \Hea\ and \Heb\ were originally reported in Refs.~\onlinecite{hir89,yip92}.
Here I highlight the electrodynamics of the BW state, as well as how that analysis led to developments in acoustic spectroscopy of superfluid \Heb, specifically the theoretical predictions and experimental discoveries of transverse sound propagation and acoustic Faraday rotation in superfluid \Heb.~\cite{moo93,lee99}

For superconductors governed by a higher dimensional representation of the spin- and orbital symmetry group the dynamics of the condensate includes Bosonic excitations beyond the $J=0^{-}$ NG (or gauge Boson) and the $J=0^{+}$ Higgs modes.
In particular, for the BW ground state the dynamics of the p-wave, spin-triplet condensate 
is governed by the non-equilibrium gap equation for the order parameter, $\vec{d}_{\vk}^{(-)}(\vq,\omega)$, 
\begin{widetext}
\be\label{eq-dminus_dynamics}
\int\frac{d\Omega_{\vk'}}{4\pi}V_{t}(\vk,\vk')\lambda(\eta',\omega)
\left\{
\left(\omega^2 -4\Delta^2 -\eta'^2\right)\vec{d}_{\vk'}^{(-)}
+4\vec{\Delta}_{\vk'}\left(\vec{\Delta}_{\vk'}\cdot\vec{d}_{\vk'}^{(-)}\right)
-
\frac{4e}{c}\eta'\left(\vv_{\vk'}\cdot\vA\right)\vec{\Delta}_{\vk'}^{(-)}
\right\}
=0
\,,
\ee
\end{widetext}
where $\eta' = \vv_{\vk'}\cdot\vq$ and $V_t = 3g_1(\hat\vk\cdot\hat\vk')$ is the pairing interaction in the spin-triplet, p-wave pairing channel. 

\begin{table}[t]
\begin{tabular}{l|r|c|l} 
$J$&$M$ 
	& $\quad\qquad t^{(J,M)}_{\alpha i}$
	& $\quad \cY_{JM}(\hat\vp)$ 
\\ 
\hline
\hline
$0$ & $0$	
	& $\frac{1}{\sqrt{3}} \delta_{\alpha i}$
	& $1$
\\ 
\hline
& $+1$	
	& $\sqrt{3}\,\epsilon_{\alpha ij}\eplus_j$
	& $-\sqrt{\frac{3}{2}}\,\hat\vp_+$
\\ 
$1$ & $0$	
	& $\sqrt{3}\,\epsilon_{\alpha ij}\e0_j$
	& $+\sqrt{3}\,\hat\vp_z $
\\ 
& $-1$	
	& $\sqrt{3}\,\epsilon_{\alpha ij}\eminus_j$
	& $+\sqrt{\frac{3}{2}}\,\hat\vp_-$
\\ 
\hline
& $+2$	
	& $\;\eplus_{\alpha}\eplus_i$
	& $+\sqrt{\frac{15}{8}}\,\hat\vp_+^2$
\\ 
& $+1$	
	& $\sqrt{\frac{1}{2}}\left(\e0_{\alpha}\eplus_i+\eplus_{\alpha}\e0_i\right)$
	& $-\sqrt{\frac{15}{2}}\,\hat\vp_z\hat\vp_+$
\\ 
$2$ & $0$	
	& $\sqrt{\frac{3}{2}}\left(\e0_{\alpha}\e0_i - \frac{1}{3}\delta_{\alpha i}\right)$
	& $+\sqrt{\frac{5}{4}}\left(3\hat\vp_z^2 - 1\right)$
\\ 
& $-1$	
	& $\sqrt{\frac{1}{2}}\left(\e0_{\alpha}\eminus_i+\eminus_{\alpha}\e0_i\right)$
	& $+\sqrt{\frac{15}{2}}\,\hat\vp_z\,\hat\vp_-$
\\ 
& $-2$	
	& $\;\eminus_{\alpha}\eminus_i$
	& $+\sqrt{\frac{15}{8}}\,\hat\vp_-^2$
\\ 
\hline
\end{tabular}
\caption{
Irreducible tensors, $\{t^{(J,M)}_{\alpha i}\}$, for $\point{SO(3)}{J}$ and $J\le 2$. 
Also included are the corresponding spherical harmonics, 
$\cY_{JM}(\hat\vp)$. 
The base unit vectors:
$\e0 = \hat\vz$, 
$\eplus = -\frac{1}{\sqrt{2}}\left(\hat\vx + i \hat\vy\right)$ 
and
$\eminus = +\frac{1}{\sqrt{2}}\left(\hat\vx - i \hat\vy\right)$,
are orthonormal: $\ve^{(\mu)*}\cdot\ve^{(\nu)}=\delta_{\mu\nu}$.
}
\label{table-tensors}
\end{table}

The solutions to the homogeneous equation ($\vA=0$) are the {\it eigenmodes} of the time-dependent equation for the Cooper pair excitations.
For $\vq=0$ the eigenvalue equation is solved by expressing the components of $\vec{d}_{\vk}$ as $d^{\alpha}_{\vk}=d_{\alpha i}\,\hat\vk_i$. Since the BW ground state is invariant under {\it joint} spin and orbital rotations, $d_{\alpha i}$ is a second rank tensor with respect to joint spin and orbital rotations.
The basis that decouples the spin-triplet, p-wave Bosonic eigenmodes are the spherical tensors, $t_{\alpha i}^{(JM)}$, corresponding to total angular momentum $J=0,1,2$ with projections, $M=0,\pm 1,\ldots,\pm J$ (c.f. Table~\ref{table-tensors}).~\cite{sau84a,sau17} The decoupling based on the total angular momentum is exact for $\vq=\mathbf{0}$. In this limit the Tsuneto function reduces to
\be\label{eq-Tsuneto_q=0}
\lambda(\omega) = 
|\Delta|^2
\int_{|\Delta|}^{+\infty}
\frac{d\varepsilon}{\sqrt{\varepsilon^2-|\Delta|^2}}
\frac{\tanh\left(\frac{\beta\varepsilon}{2}\right)}
     {\varepsilon^2-(\omega+i\gamma)^2/4}
\,,
\ee
where $\gamma\rightarrow 0^+$ ensures the causal (retarded) response of the condensate to the external field. For frequencies below the pair-breaking threshold, $\hbar\omega<2|\Delta|$, the Tsuneto function is real and positive, and provides the stiffness of the condensate in response to an  external field. However, for photon (phonon) energies $\hbar\omega\ge 2|\Delta|$ single photon (phonon) absorption leads to dissociation of Cooper pairs into unbound pairs of quasiparticles. 
The Tsuneto function acquires an imaginary part corresponding to the spectral density 
of unbound pairs,
\be
\Im\lambda(\omega) = 2\pi\frac{\tanh\left(\displaystyle{\frac{\omega}{4T}}\right)}{\omega}
                     \frac{|\Delta|^2}{\sqrt{\omega^2-4|\Delta|^2}}
\,,\omega>2|\Delta|
\,.
\ee
See Ref.~\onlinecite{mck90} for more discussion and evaluation of the Tsuneto function $\lambda(\omega)$.

The linear coupling to the gauge field $\vA$ allows {\it only} Bosonic modes with $J=0$, $J=2$ and $\charge=-1$ to be excited by the EM field. For these modes the eigenvectors and corresponding momentum-space eigenfunctions are given in Table~\ref{table-tensors}, while the corresponding mass (eigenfrequency) and quantum numbers for the Bosonic modes of the BW ground state are given in Table~\ref{table-BW_modes}.
The couplings of these modes to the EM field is determined by the polarization state of the gauge field, $\vA$, and the direction of propagation, $\vq$. 

\vspace{-5mm}
\subsection{Dynamics of NG and Higgs Amplitudes}
\vspace{-3mm}

To determine the contributions of the Bosonic modes to the charge current I solve the dynamical equations by expanding $d_{\alpha i}$ in the basis tensors, $t_{\alpha i}^{(JM)}$, with the quantization axis for the modes chosen to be $\vq$,
\be
d^{(-)}_{\alpha i}=\sum_{JM}\cD^{(-)}_{JM}(\vq,\omega)\,t^{(JM)}_{\alpha i} 
\,.
\ee
The spherical tensors satisfy the orthogonality and normalization conditions,
\be
\Tr{\widehat{t}^{(JM)}\widehat{t}^{(J'M')\dag}}
=
\delta_{JJ'}\,\delta_{MM'}
\,,
\ee
which are used to project out the Bosonic mode amplitudes $\cD^{(-)}_{JM}(\vq,\omega)$ from 
Eq.~\ref{eq-dminus_dynamics}. 
Note that the anti-symmetric Bosonic modes with $J=1$ do not couple to the gauge field. 
Similarly, the modes with $J=2^{-}$, $M=\pm 2$ do not couple to the gauge field.
As noted earlier the phase mode $\cD^{(-)}_{00}$, and by extension the $J=2$, $M=0$, $\charge=-1$ mode, can be absorbed into the gauge field by fixing the gauge of $\vA$. However, in the gauge-invariant formulation, $\cD^{(-)}_{00}$ and $\cD^{(-)}_{20}$ modes couple only to the scalar potential and longitudinal component of $\vA$.
For transverse fields, $\vq\cdot\vA = 0$, only the Bosonic modes with $J=2, M=\pm 1$ and $\charge=-1$ are excited by the EM field, 
\be
\cD^{(-)}_{2,\pm 1} 
=
\frac{4\Delta}{5}\left(\frac{2e}{c}\right)\,
\frac{v_f^2\left[\vq_{\alpha}t^{(2,\pm 1)*}_{\alpha i}\,\vA_i\right]}
{\left[(\omega+i\gamma)^2 - \Omega_{2,\pm 1}(q)^2 \right]}
\,.
\label{eq-D2pm1}
\ee
All other matrix elements, $\vq_{\alpha} t^{(JM)}_{\alpha i} \vA_i$, vanish. Thus, the $J=2$, $M=\pm 1$ modes are resonantly excited by a transverse EM field for frequencies tuned to the eigenfrequencies,
\be\label{eq-mode-dispersion}
\Omega_{2,\pm 1}(q)=\sqrt{\Omega_{0}^{2}+c^{2}_{2,1}q^{2}}
\,,
\ee
where $\Omega_{0}=\sqrt{\frac{12}{5}}\,\Delta$ and $c_{2,1}=\sqrt{\frac{2}{5}}\,v_f$ in the absence of vacuum polarization corrections and Fermi surface anisotropy.~\cite{sau17,sau15} The dispersion of the modes plays a significant role in the EM power spectrum as I discuss below.

Impurity scattering leads to pair breaking which modifies the collective mode response 
by reducing the mass of the Bosonic mode as well as generating sub-gap quasiparticle excitations that lead to a finite lifetime of the mode. 
A detailed theory of the impact of disorder on the collective mode spectrum is outside the scope of this report, but I include the finite lifetime of the modes phenomenologically by replacing $\omega\rightarrow\omega+i\gamma$, where $\gamma=1/\tau>0$ is the inverse of the mode lifetime and is of order the mean collision rate for electron-impurity scattering. In what follows I assume $1/\tau\ll\Delta$.
Based on the form of Eqs.~\ref{eq-Current-response} and \ref{eq-Gk} the current response can be written in the form
\be
\vJ_i(\vq,\omega)
=
-\left[\mathbb{K}_{ij}^{\text{QP}}+\mathbb{K}^{\text{CM}}_{ij}\right]\vA_{j}
\,,
\ee
where $\mathbb{K}^{\text{QP}}$ is the current response of the quasiparticle spectrum, while $\mathbb{K}^{\text{CM}}$ gives the current response from the collective modes with $J=2$, $M=\pm 1$, $\charge=-1$.
The isotropy of the quasiparticle spectrum for the BW ground state implies the quasiparticle current response function is given by $\mathbb{K}_{ij}^{\text{QP}}=K^{\text{QP}}\delta_{ij}+\mathbb{K}_{||}^{\text{QP}}\hat\vq_i\hat\vq_j$, with
\begin{widetext}
\be
K^{\text{QP}}
=
\left(\frac{ne^{2}}{m^*c}\right)
\left\lbrace
1+\threehalves
\int\frac{d\Omega_{\hat\vk}}{4\pi}
\left(1 - (\hat\vq\cdot\hat\vk)^2\right)
\left[1 - \lambda(\vv_{\vk}\cdot\vq,\omega)\right]
\left(
\frac{\vv_{\vk}\cdot\vq}{\omega+i\gamma - \vv_{\vk}\cdot\vq}
\right)
\right\rbrace
\,.
\ee
\end{widetext}
For transverse fields $\mathbb{K}^{\text{QP}}_{||}$ does not contribute to $\vJ(\vq,\omega)$. For temperatures above $T_c$, $K^{\text{QP}}$ is proportional to the frequency dependent conductivity for an ultra-clean metal ($\gamma\rightarrow 0^{+}$), and determines the power absorption in the \emph{anomalous skin regime},
\be
K^{\text{QP}}_{\text{N}}
\equiv
-\frac{i\omega}{c}\,\sigma_{\text{N}}(q,\omega)
\simeq 
-i\pi\,\frac{3}{4}
\left(\frac{ne^{2}}{m^*c}\right)\,\frac{\omega}{v_fq}
\,.
\ee
The contribution to the current response from the $J=2, M=\pm 1$ collective modes reduces to, $\mathbb{K}_{ij}^{\text{CM}}=K^{\text{CM}}(\delta_{ij}-\hat\vq_i\hat\vq_j)$, and in the same ultra-clean limit reduces to,
\be
K^{\text{CM}}
= 
\frac{6}{5}
\left(\frac{ne^2}{m^*c}\right)
\frac{(v_fq)^2\,I(q,\omega)}
     {\left[(\omega+i\gamma)^2 - \Omega_{2,1}(q)^2\right]}
\,,
\ee
\be
I(q,\omega)
\ns=\ns
\onehalf\ns\int\ns\frac{d\Omega_{\hat\vk}}{4\pi}
(\hat\vk\cdot\hat\vq)^2\left[1-(\hat\vk\cdot\hat\vq)^2\right]
\lambda(\vv_{\vk}\cdot\vq,\omega)
\,.
\ee

\begin{figure}[t]
\includegraphics[width=0.95\linewidth,keepaspectratio]{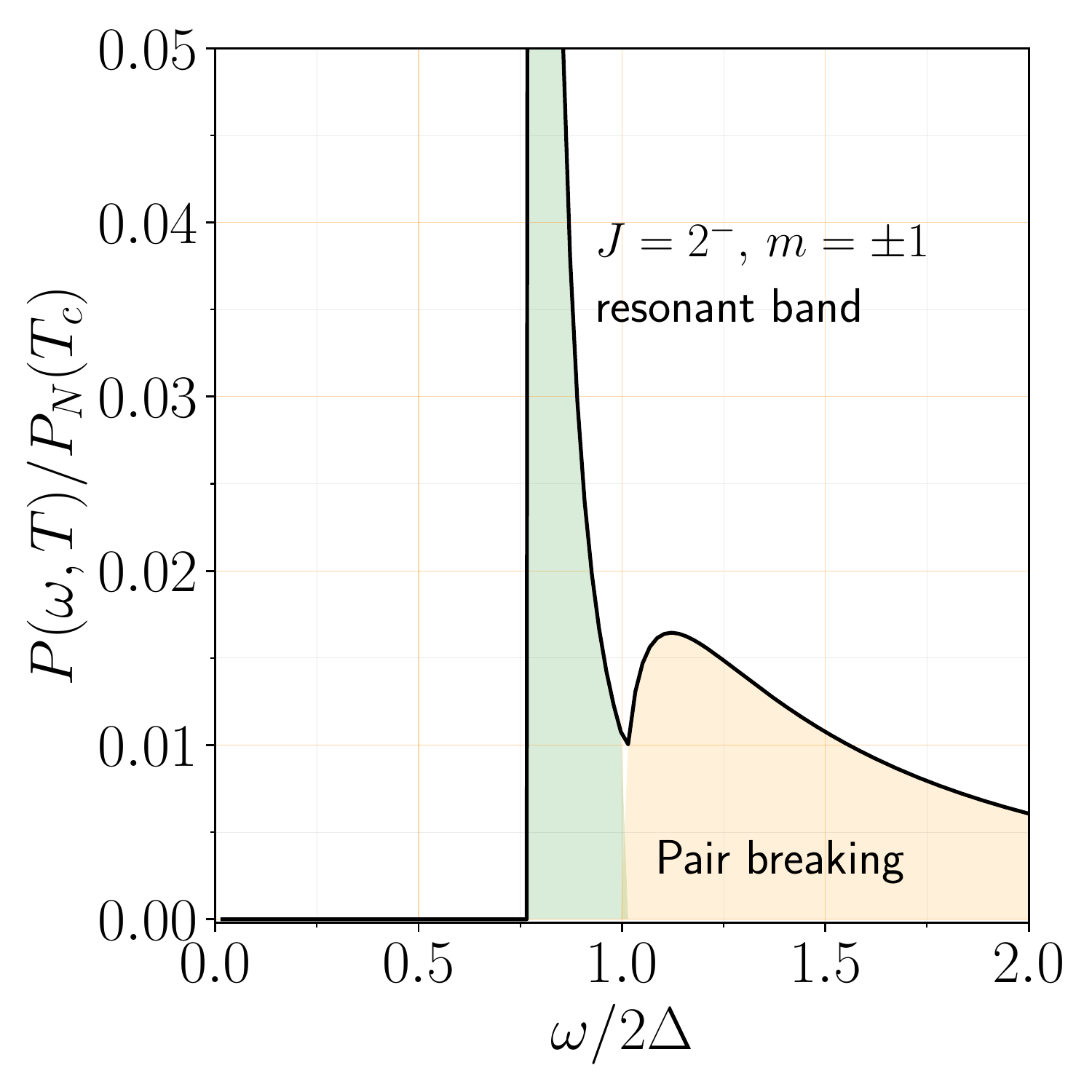}
\caption{\small
Power absorption spectrum of EM radiation for the Balian-Werthamer state at $T=0$ normalized to the power absorption of the normal state in the anomalous skin regime, $P_{\text{N}}(\omega)$, for $\Lambda/\xi_0=10.0$. 
\label{fig-power_absorption}
}
\end{figure}

\vspace{-5mm}
\subsection{Power Absorption Spectrum}
\vspace{-3mm}

The power absorption spectrum is obtained by integrating the Joule dissipation at frequency $\omega$ over the half-space of the metal,
\be
P(\omega)=\frac{1}{2}\int_0^{\infty}\,dz\,
           \Re\ns\left\lbrace\vE_{\omega}^{*}(z)\cdot\vJ_{\omega}(z)\right\rbrace
\,.
\ee
At the vacuum-metal interface the EM field penetrates a distance $z$ of order the skin depth into the metal, or the London penetration depth in the superconducting state. 
Below I calculate the contributions from quasiparticles and the $J=2^{-}, M=\pm 1$ Higgs modes to the power absorption spectrum.
I consider specular boundary conditions for the vacuum-superconducting interface, and omit the effect of surface pair-breaking, valid in the strong type II limit $\Lambda\gg\xi$. 
%
The power absorption is then dominated by bulk quasiparticles and the $J=2^{-}$, $M=\pm 1$ Higgs modes with wavevectors in the range $q \lesssim 1/\Lambda\ll 1/\xi$.
The half-space boundary value problem is mapped onto a full-space boundary value problem 
for specular boundary conditions at the vacuum-metal interface.
Specifying the magnetic field strength, $B_0(\omega)$, at the vacuum side of the interface, we obtain from Maxwell's equations and the continuity of $\vA$ at the boundary (c.f. p. 373 of Ref.~\onlinecite{LL10}),
\be
\vA(\vq,\omega)=\frac{2B_0(\omega)}{q^{2}+\frac{4\pi}{c}\,K(\vq,\omega)}\,\hat\ve
\,,
\ee
where $\hat\ve\perp\vq$ is the polarization direction of the transverse EM field.
The result for the power absorption then becomes,
\be
P(\omega)=\frac{2\omega}{c} \left\vert B_0(\omega)\right\vert^2
\int_0^{\infty}\frac{dq}{2\pi}\,
\frac{\Im K(\vq,\omega)}
     {\left\vert q^2+\frac{4\pi}{c}\,K(\vq,\omega) \right\vert^2}
\,.
\ee
As a basis for comparison the normal-metal power absorption in the anomalous skin limit is,
\be
P_{\text{N}}(\omega)=\frac{1}{8\sqrt{3}}
\left(\frac{2}{3\pi}\right)^{\frac{4}{3}}
\left\vert B_0(\omega)\right\vert^2\,\left(\omega^2\Lambda^2 v_f\right)^{\nicefrac{1}{3}}
\,,
\ee
where, $\Lambda=c/\omega_p=\sqrt{m^* c^2/4\pi ne^2}$ is the zero-temperature London penetration depth.

In the superconducting state the power absorption from the $J=2^{-},M=\pm 1$ modes is given by,
\be
P_{\text{CM}}(\omega)=\frac{2\omega}{c}\left\vert{B_0}(\omega)\right\vert^{2} 
\int_0^{\infty}\frac{dq}{2\pi}\,
\frac{\Im K^{\text{CM}}}{\left\vert{q^{2}+\frac{4\pi}{c}K(q,\omega)}\right\vert^2}
\,,
\ee
\be
\hspace*{-3mm}
\Im K_{\text{CM}}\simeq\frac{2\pi}{25}\left(\frac{ne^2}{m^*c}\right)(v_fq)^2
\lambda(\omega)\delta(\omega^2-\Omega_{2,1}(q)^2)
\,,
\ee
where $\lambda(\omega)\equiv\lambda(0,\omega)$. It is convenient to introduce the wave vector, $q_1(\omega)=\Theta(\omega-\Omega_0)\sqrt{\omega^2-\Omega_0^2}/c_{2,1}$, corresponding to the resonance condition $\omega\equiv\Omega_{2,1}(q_1)$, where $\Omega_0=\sqrt{12/5}\Delta$ is the threshold frequency and $c_{2,1}=\sqrt{2/5}v_f$ is the velocity that determines the dispersion of the $J=2^{-},M=\pm 1$ modes. In the limit, $q v_f\ll\omega$, the denominator of $P_{\text{CM}}(\omega)$ is to good approximation given by $K\approx(ne^2/m^*c)$, with the result,
\be
\hspace*{-2mm}
P_{\text{CM}}(\omega)
\ns=\ns
\frac{\left\vert B_{0}(\omega) \right\vert^2}{100\pi} 
\ns
\left|\frac{v_f}{c_{2,1}}\right|^2\ns\omega\Lambda\lambda(\omega)
\frac{q_1(\omega)\Lambda}{\left[1 + (q_{1}(\omega)\Lambda)^2\right]^2}
\,,
\ee
for $\omega<2\Delta$.
At $T=0$ the power absorption in the BW superconductor vanishes for $\omega < \Omega_0$. Below this threshold only the supercurrent is excited by the EM field. Absorption of EM radiation onsets and increases rapidly for $\omega>\Omega_0$, reaching a maximum at the frequency,
\be
\Omega_{*}\approx\Omega_{0}+\frac{1}{6}\frac{c_{2,1}^2}{\Lambda^2\Omega_0} < 2\Delta\,,
\,\mbox{for}\,\Lambda\gg\xi_{\text{$\Delta$}}=\frac{\hbar v_f}{\pi\Delta}
\,.
\ee
Above the continuum edge ($\hbar\omega > 2\Delta$) quasiparticle excitations produced by dissociation of Cooper pairs contribute to the power absorption. The full power absorption spectrum resulting from both pair dissociation and resonant excitation of the $J=2$, $M=\pm 1$ collective modes is shown in Fig.~\ref{fig-power_absorption}.
Note the onset at $\omega=\Omega_0\approx 1.55\Delta$ and the peak absorption below the continuum edge at $\Omega_*$. The peak absorption at $T=0$ normalized to the absorption in the normal state at $\Omega_*$ is 
\be
\frac{P_{*}}{P_{N}}\ns=\ns\frac{9(\frac{3\pi}{2})^{\frac{4}{3}}}{200\pi}
            \ns\left|\frac{v_f}{c_{2,1}}\right|^2\lambda(\Omega_*)
           \ns\left|\frac{\Omega_*\Lambda}{v_f}\right|^{\frac{1}{3}}
            \ns\simeq 0.28\,\left|\frac{\Omega_*\Lambda}{v_f}\right|^{\frac{1}{3}}
\,,
\ee
which is easily of order 1, and can be substantially larger for 
$\Lambda\gg\xi_{\text{$\Delta$}}$, as shown in Fig.~\ref{fig-power_absorption} for $\Lambda/\xi_0=10$.

Thus, just as acoustic spectroscopy provided confirmation of the B phase of \He\ as the BW state, such an EM power absorption spectrum would provide direct evidence of an electronic realization of the BW state.
It is also worth noting that this theoretical prediction for the excitation of the $J=2^{-}$, $M=\pm 1$ modes 
by a transverse EM field was the seed that led to the prediction of propagating transverse sound in \Heb\ by G. Moores and me in 1993.~\cite{moo93} The basic idea was that transverse currents in the neutral BW phase would become de-confined from the surface since the London screening length diverges as $e\rightarrow 0$. 
The full theory requires a detailed analysis of the restoring forces for propagating transverse mass currents, but the basic idea holds with the $J=2^{-}$, $M=\pm 1$ Higgs modes amplifying the restoring force for transverse zero sound at frequencies $\omega > \Omega_{2,\pm 1}\simeq \sqrt{12/5}\Delta(T)$.~\cite{moo93}
Nuclear Zeeman splitting of the $J=2^{-}$, $M=\pm 1$ modes in a magnetic field $\vH\parallel\vq$ leads to circular birefringence of right- and left-circularly polarized transverse mass currents. Thus, a linearly polarized transverse wave undergoes Faraday rotation.~\cite{moo93,sau00b} 

The experimental observation of transverse sound in \Heb\ at ultra-sound frequencies, $\omega/2\pi\approx 82\,\mbox{MHz}$, followed the theoretical prediction, with direct observation of the transverse nature of the propagating mode provided by measurements of Faraday rotation of the mass current polarization for \Heb\ in a static magnetic field along
the propagation direction of the mode, $\vq\parallel\vH$.~\cite{lee99}
This remarkable discovery revealed emergent physics: application of a magnetic field that couples to the nulcear magnetic moment of the $J=2^{-}$, $M=\pm 1$ Cooper pairs generates a torque that rotates the direction of the mass current of transverse sound!
The stiffness that transmits the torque derives from the spontaneous breaking of \emph{relative spin-orbit rotation symmetry}. Acoustic Faraday rotation is perhaps the most direct observation of this subtle broken relative symmetry.

\begin{figure}[t]
\includegraphics[width=0.95\linewidth,keepaspectratio]{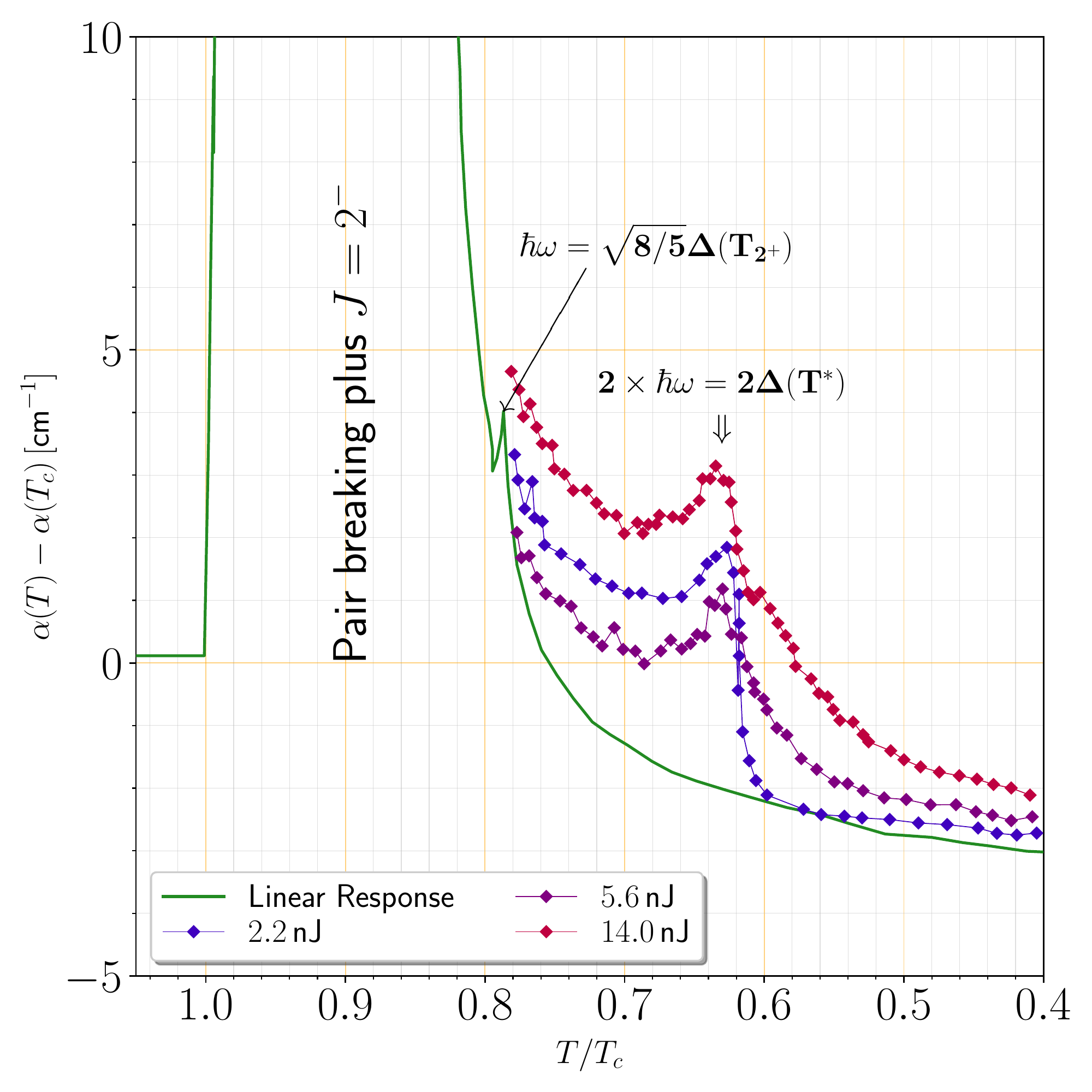}
\caption{\small
Appearance of a feature in the absorption spectrum of zero sound in \Heb\ at frequency $\omega/2\pi= 50\,\mbox{MHz}$ and a pressure of $p=5.3\,\mbox{bar}$ onsetting sharply at $T_*=0.63\,T_c$ corresponding to $2\times\hbar\omega = 2\Delta(T_*)$. The attenuation peak 
is observed at high powers: 
$2.2\,\mbox{nJ}$, 
$5.6\,\mbox{nJ}$, 
and $14\,\mbox{nJ}$, and is identified with two-phonon excitation of the $J=0^+$ Higgs mode. 
The excess attenuation at temperatures $T>T_*$
is identified with two-phonon pair-breaking.
The thin line is the expected attenuation in the linear response limit. 
The figure is plotted from the data reported by Peters and Eska.~\cite{pet92} 
}
\label{fig-two_phonon_absorption}
\end{figure}

\vspace{-5mm}
\subsection{The $J=0^+$ Higgs mode}\label{sec-J=0+_Higgs}
\vspace{-3mm}

In conventional superconductors the amplitude Higgs mode has the same quantum numbers ($L=0$, $S=0$, $\charge=+1$) as the condensate vacuum. 
As a result excitation of the Higgs mode by single photon (phonon) absorption is forbidden. This also applies to the $J=0^{+}$ Higgs mode of the BW state.
In fact if $\charge$ parity is an exact symmetry of the parent Fermionic vacuum, then single photon (phonon) transitions from the ground state to any $J^{+}$ mode is forbidden.

The selection rule can be avoided by either a two-photon (two-phonon) coupling to the Higgs mode, or lifted by explicitly breaking particle-hole symmetry. Thus, inelastic photon (phonon) scattering (Raman scattering) can excite the $J=0^{+}$ Higgs mode, or two-photon (two-phonon) absorption with resonant excitation of the $J=0^{+}$ Higgs mode is an allowed process.
Indeed the first observation of the amplitude Higgs mode was in superconducting \nbse, in which a peak in the Raman scattering spectrum for energy transfer near $\hbar\omega=2\Delta$ that develops below the superconducting transition.~\cite{soo80}$^{,}$\footnote{Note that the discoveries of the $J=0^{+}$ Higgs mode in \nbse\ and the $J=2^{+}$ Higgs mode in \Heb\ were all published in Physical Review Letters within a month of each other in the summer of 1980.}
See also more recent experiments in zero field.~\cite{mea13}
The theory of the coupling to the Higgs mode in \nbse, for which the charge density wave phonon plays as key role, was worked out by Littlewood and Varma.~\cite{lit81} 

Ultrasound propagation and absorption experiments at high power by Peters and Eska in \Heb\ revealed two-phonon pair breaking for excitation energies $2\,\hbar\omega\ge 2\Delta(T)$, as shown in Fig.~\ref{fig-two_phonon_absorption}.~\cite{pet92}
At zero temperature single phonon pairbreaking onsets as a threshold with $\alpha(\omega)\sim\sqrt{\omega-2\Delta}$, with no peak near $2\Delta$, even though the pair-breaking spectral density scales as $\Im\lambda(\omega)\sim 1/\sqrt{\omega-2\Delta}$. 
Assuming two-phonon pairbreaking onsets similarly then the excess attenuation that onsets sharply at $T_*=0.63\,T_c$ corresponding to $2\times\hbar\omega = 2\Delta(T_*)$ \emph{with a peak} on the leading edge suggests two-phonon excitation of the $J=0^{+}$ Higgs mode was observed in \Heb\ in 1992. Theoretical predictions for the structure of the two-phonon absorption edge are needed in order to provide a definitive interpretation of the two-phonon absorption peak.

Koch and W\"olfle introduced the mechanism of particle-hole asymmetry to lift the $\charge$ parity selection rule. Their mechanism leads to a very small particle-hole asymmetry parameter.   
However, tuneable particle-hole asymmetry is also possible. In particular, a supercurrent lifts the $\charge$ parity selection rule for single phonon transitions to any of the $J^{+}$ modes with an asymmetry parameter proportional to $v_s/v_c$, where $v_s$ is the condensate velocity and $v_c=\Delta/p_f$ is the bulk critical velocity.
Indeed the theory of current-induced coupling to the $J^{+}$ Higgs spectrum was pioneered in the context of \Heb\ with the prediction\cite{sau84a,sau91,mck90} and discovery\cite{tor92,man94} 
of parametric excitation of the $J=2^{+}$ modes.
This mechanism of parametric excitation of the $J=0^{+}$ Higgs mode has been sucessful in conventional superconductors~\cite{mat13}, and has opened new directions in nonequilibrium superconductivity.~\cite{vas21}  

\vspace{-7mm}
\section{Summary \& Outlook}
\vspace{-3mm}

Bosonic excitations of a Cooper pair condensate are generic features of superconductors. For conventional spin-singlet, s-wave BCS superconductors these Bosonic excitation are the massless Anderson-Bogoliubov phase mode and the amplitude Higgs mode. The Anderson-Bogoliubov mode is the Nambu-Goldstone Boson associated with broken $\gauge$ symmetry and is observable as collisionless sound in neutral BCS superfluids, while the amplitude Higgs mode is elusive, difficult to excite since it has the same quantum numbers as the condensate vacuum, and difficult to distinguish since its mass conicides with the threshold for dissociation of Cooper pairs.

For electrically charged superconductors the condensate phase can be absorbed into the gauge field with a gauge fixing condition. The remaining dynamics describes a gauge Boson obeying a Klein-Gordon field equation with a mass $M_{\text{A}}$ and wavelength, $\Lambda = \hbar/M_{\text{A}}c$, corresponding to the London penetration depth. Thus, the primary collective mode response in conventional superconductors is that of persistent currents responsible for the Meissner screening and flux confinement in quantized vortices.

For superconductors that break additional symmetries in conjunction with $\gauge$ symmetry, and belong to a multi-dimensional representation of the maximal symmetry group of the normal metallic state, additional collective modes emerge that contribute to the electrodynamics, acoustics or hydrodynamics of the pair condensate. 
In superfluid \Heb, the realization of the BW state, these order parameter collective modes have been studied extensively in relation to the propagation and attenuation of ultrasound at frequencies $\omega\approx\Delta/\hbar\approx 50-100\,\mbox{MHz}$.
However, the corresponding role of collective modes in charged unconventional superconductors is comparatively unexplored experimentally.
The realization of a Balian-Werthamer superconductor would exhibit a novel electrodynamics, including EM absorption signatures of the sub-gap Higgs modes at GHz to THz frequencies, ESR signatures to d.c. signatures of surface Majorana modes in the London screening current. 

\vspace*{-5mm}
\subsection*{Acknowledgements}
\vspace*{-3mm}

This work was supported by National Science Foundation Grant DMR-150873.
I thank Anton Vorontsov for his translation of the Vdovin paper to English, and for discussions on the analysis in Vdovin's paper. 
I thank Bill Halperin for many discussions on the remarkable phenomena revealed by acoustic, NMR and thermodynamic studies in pursuit of understanding the BW state of \Heb.
I congratulate Dave Lee and John Reppy on the occasion of their 90th birthdays and especially thank them for all the beautiful physics they have revealed by pushing the frontiers of quantum fluids and solids at ultra-low temperatures.  

\vspace{-3mm}
\appendix
\vspace{-5mm}

\section{Vdovin's Contribution}\label{sec-Vdovin}
\vspace{-3mm}

In 1987 I gave an invited talk at the March meeting of the American Physical Society held in New York on the ``Theory of Sound Propagation and Attenuation in Superfluid $^3$He'' including the Zeeman and Paschen-Back effects of the $J=2^{\pm}$ collective modes. 
This was the ``Woodstock of Physics'' meeting that highlighted the discovery of high-temperature superconductivity in the cuprates.
Tony Leggett was the chair of my session and after the talk he told me that the collective mode spectrum had been obtained by Yu. Vdovin years before the discovery of \Heb\ and the theoretical works of W\"olfle, Serene and Maki on the collective modes and their acoustic signatures. Tony kindly sent me a copy of a collection of articles published in Moscow in 1963 on ``Methods of Quantum Field Theory to the Many Body Problem'', which included the article by Vdovin titled ``Effects of pairing in Fermi systems in a P-state''.~\cite{vdo63} Tony also drew my attention to a sentence at the end of the abstract stating that the work had been completed in 1961! That was two years before the publication of the work by Balian and Werthamer, and the same year as the publication of the papers by Anderson and Morel,\cite{and61} Gorkov and Galitskii,\cite{gor61} and Vaks, Galitskii and Larkin.\cite{vak61} 
As far as I know the first reference to Vdovin's paper in the literature on \He\ or collective modes in superconductors was my review of collective modes and nonlinear acoustics with R. McKenzie in 1990.\cite{mck90} 
About the same time Vollhardt and W\"olfle cited Vdovin's paper in their treatise, ``The Superfluid Phases of Helium 3'', and pointed out that Vdovin's work ``fell into oblivion''. That appears to be true, as Vaks, Galitskii and Larkin, who published work on collective excitations in higher angular momentum states in 1962,\cite{vak62} appear to have been unaware of Vdovin's work.
However, the connection between Vdovin's paper and these four early papers on the theory of pairing in higher angular momentum states is I think worth clarifying in an article reflecting on the impact of the BW ground state on both the physics of superfluid \He, as well as the theory of unconventional superconductors.
The existence of Vdovin's early work, and that it appears to have been done as early as 1961, has been interpreted to imply that Vdovin should be credited equally with Balian and Werthamer for the theoretical prediction for the ground state of a spin-triplet, p-wave superconductor, i.e. what I have referred to as the BW ground state, c.f. Ref.~\onlinecite{hal19}. However, that is incorrect.

Balian and Werthamer \emph{proved} that the $^3$P$_0$ state with $L=1$, $S=1$ and $J=0$ was the absolute minimum of the weak-coupling BCS free energy functional within the p-wave/spin-triplet manifold. See Sec.~3, p.~1556 of the BW paper.~\cite{bal63} The physical reason is that within the most attractive pairing channel the lowest free energy state(s) is the linear superposition that maximizes the pairing gap over the Fermi surface, and for the spin-triplet, p-wave manifold this is the BW state.~\footnote{The BW state remains the ground state within weak coupling theory even with an additional attractive, but subdominant, pairing channel, e.g. an attractive f-wave pairing interaction.\cite{sau86} However, as is well known, strong-coupling corrections to the weak-coupling free energy functional stabilize anisotropic states. Indeed the A phase is the realization of the anisotropic Anderson-Morel state.}

Vdovin made no such analysis of the stability of phases within the p-wave, triplet manifold. Rather he assumed the ground state was the $^3$P$_0$ state. From paragraph 3 on p. 95 of Ref.~\onlinecite{vdo63}, ``Both single-particle and collective excitations are considered in this system. Different branches of the collective excitation spectrum, corresponding to dynamics of bound pairs with different moments J, are obtained in the assumption that the condensate is made from pairs in $^3$P$_0$ state.'' 

The basis for Vdovin's assumption of a $^3$P$_0$ ground state is the paper of Gorkov and Galitskii (GG).~\cite{gor61} However, the paper by GG contains fundamental errors and is not a proof that the $^3$P$_0$ state is the ground state. GG start from an \emph{ansatz} for the two-particle density matrix, $\rho_{\alpha\beta;\gamma\rho}(p,-p;p',-p')\equiv\langle\psi_{\alpha}(p)\psi_{\beta}(-p)\psi^{\dag}_{\gamma}(p') \psi^{\dag}_{\delta}(-p')\rangle$, which is not a BCS condensate, but rather a fragmented condensate,~\cite{noz95} i.e. $(2l+1)$ condensates with macroscopic eigenvalues of the form (1st equation on p.~793 of Ref.~\onlinecite{gor61}),
\be
\rho_{\alpha\beta;\gamma\rho}(p,-p;p',-p')
\rightarrow
\sum_{m=-l}^{+l}
F_{m,\alpha\beta}(p)\,F^{\dag}_{m,\gamma\delta}(p')
\,.
\ee
GG posit an equation for each $m$ of the form,
\be\label{eq-GG_Delta_m}
\hat{\Delta}_m(\vp) = \int d\vp'\,V(|\vp-\vp'|)\,\int d\omega\,\hat{F}_m(\vp',\omega)
\,,
\ee
then assert that ``since the angular momentum is zero'', the diagonal (quasiparticle) propagator is isotropic
with 
\be
G_{\alpha\beta}(p) = G(|\vp|,\omega)\,\delta_{\alpha\beta}
\,.
\ee
With this assumption GG eliminate all pairing states that do not have an isotropic excitation gap. Specifically GG argue that since the diagonal propagator is isotropic then $\hat{F}_m(p)\propto Y_{lm}(\vp)$ as is $\hat{\Delta}_m(p)$, and thus based on Eq.~\eqref{eq-GG_Delta_m}, each $\hat{\Delta}_m(p)$ has the same amplitude, in which case the addition theorem for the spherical harmonics generates an isotropic excitation gap given by,
\be
|\Delta|^2 = \frac{1}{2}|\Delta_m|^2\,(2L+1)(2S+1)\,P_{L}(\theta=0)
\,.
\ee
It is a circular argument disconnected from the BCS free energy functional and the BCS gap equation, which is the stationarity condition of the former.~\cite{sau86}

By contrast BCS condensation corresponds to macroscopic occupation of a single two-particle state
\be
\rho_{\alpha\beta;\gamma\rho}(p,-p;p',-p')
\rightarrow
F_{\alpha\beta}(p)\,F^{\dag}_{\gamma\delta}(p')
\,,
\ee
where the spin- and orbital structure of the Cooper pair amplitude, 
$F_{\alpha\beta}(p)=\langle\psi_{\alpha}(p)\psi_{\beta}(-p)\rangle$ 
is determined self-consistently by the BCS mean field gap equation,
\be
\hat{\Delta}(\vp) = \int d\vp'\,V(|\vp-\vp'|)\,\int d\omega\,\hat{F}(\vp',\omega)
\,.
\ee
The linearized form of the gap equation separates into a set of eigenvalue equations determined by pairing interactions, $V_l$, for each of the irreducible representations of the symmetry group of the normal state, which in this case is $\point{SO(3)}{L}$. The superconducting transition is then driven by the most 
attractive pairing interaction, e.g. $V_1$, resulting in an anomalous self energy of the form
\be
\hat\Delta(\vp) = \sum_{m_s=-1}^{+1}\sum_{m_L=-1}^{+1}\Delta_{m_s,m_L}\,\hat{S}_{1,m_s}Y_{1,m}(\vp)
\,,
\ee
where $\hat{S}_{1,m_s}$ are the $2\times 2$ matrix represention of spin states $\ket{1,m_s}$ and $Y_{1,m_L}(\vp)$ are the p-wave orbital basis states, i.e. the $L=1$ spherical harmonics.
The amplitudes $\Delta_{m_s,m_L}$ are determined by solutions to the full nonlinear BCS gap equation, which is the stationarity condition for the weak coupling BCS free energy functional. The lowest energy state among the solutions to the gap equation is the ground state, which for $L=1$, $S=1$, is the BW state.

To summarize, Vdovin's contribution was the original prediction of the Bosonic collective modes based on the assumed BW ground state using the field theory method developed by Vaks, Galitskii and Larkin.\cite{vak61} However, neither Vdovin, nor Gorkov and Galitskii, proved that the ground state of a spin-triplet, p-wave superconductor is the $^3$P$_0$ state. That was the work of Balian and Werthamer.

\vspace*{-5mm}
\section{Evaluation of the response function}
\vspace*{-3mm}

\begin{figure}[t]
\includegraphics[width=0.95\linewidth,keepaspectratio]{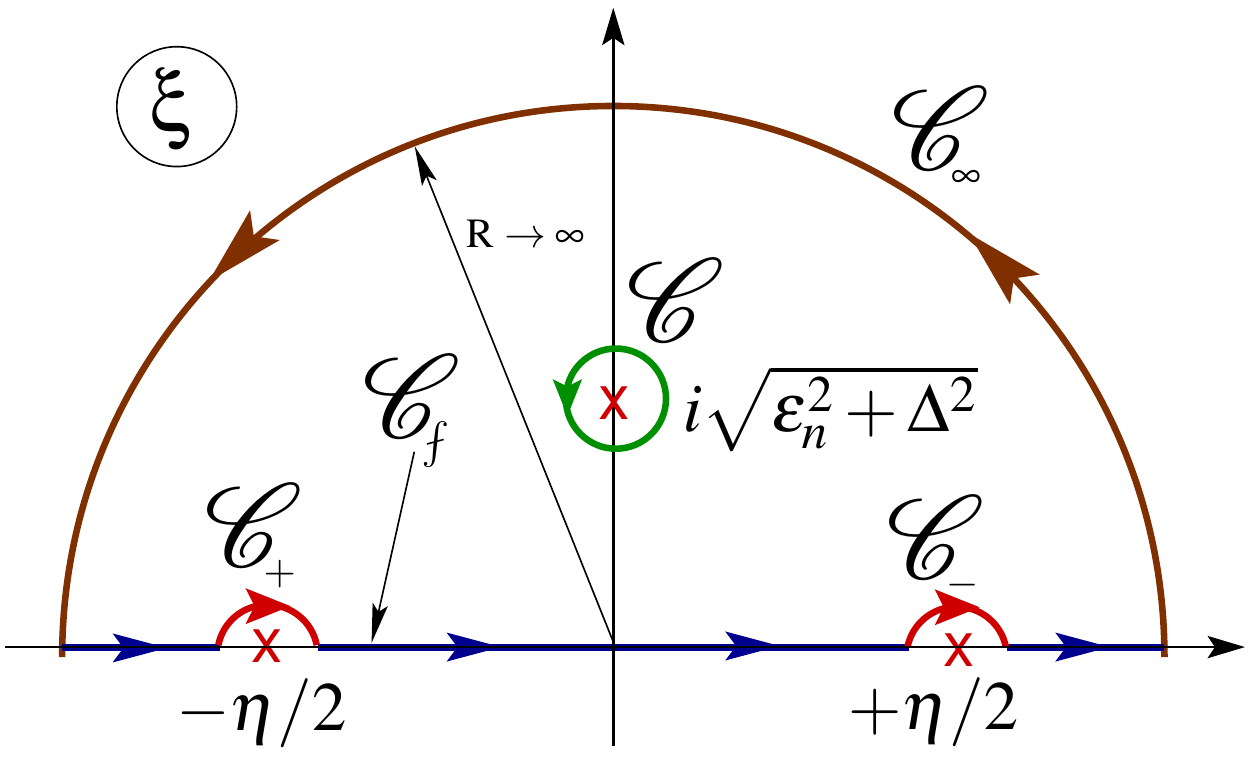}
\caption{\small
Integration contours for evaluating the principal part integral in Eq.~\ref{eq-Tsuneto_omega=0-Matsubara} for the static condensate response function $\lambda(\eta)$.
\label{fig-contour}
}
\end{figure}

Equation~\ref{eq-Tsuneto_omega=0} for the static condensate response is obtained by evaluating Eq.~\ref{eq-tsuneto_function} with $\omega=0$ and changing the integration variable to $\xi=\sgn(\varepsilon)\sqrt{\varepsilon^2-|\Delta|^2}$. The symbol $\fint$ implies principal part integration in the neighborhood of the singularities on the real $\xi$ axis at $\pm\eta/2$.
This integral is most easily evaluated by using the Matsubara representation for the hyperbolic tangent function,
\be
\frac{\tanh(\sqrt{\xi^2+|\Delta|^2}/2T)}{2\sqrt{\xi^2+|\Delta|^2}}
= 
T\sum_{\varepsilon_n}\frac{1}{\xi^2 + \varepsilon_n^2 + |\Delta|^2}
\,,
\ee
where $\varepsilon_n=(2n+1)\pi T$ are the Fermion Matsubara frequencies with $n\in\mathbb{Z}$. Thus, Eq.~\ref{eq-Tsuneto_omega=0} becomes
\be\label{eq-Tsuneto_omega=0-Matsubara}
\lambda(\eta)\ns=\ns
                 T\sum_{\varepsilon_n}
                 \fint_{-\infty}^{+\infty}\ns\ns\ns\ns d\xi
                 \frac{|\Delta|^2}{\xi^2+\varepsilon_n^2+|\Delta|^2}
                 \ns\times\ns\frac{1}{\nicefrac{1}{4}\eta^2 - \xi^2}
\,.
\ee
The principal part integral on the real axis is a component of the integral over the closed contour shown in Fig.~\ref{fig-contour}, i.e. $\cC_{\fint} + \cC_{+} + \cC_{-} + \cC_{\infty} = \cC$, where $\cC_{\fint}$ is the path of the principal part integral on the real $\xi$-axis, $\cC_{\pm}$ is an infinitesimal half circle in the upper half $\xi$-plane of radius $\delta\rightarrow 0^+$ centered at $\xi_{\pm}=\pm\eta/2$, and $\cC_{\infty}$ is a half circle in the upper half plane of radius $R\rightarrow\infty$. The integrand 
\be
I(\xi)=\frac{1}{\xi^2+\varepsilon_n^2+|\Delta|^2}
       \times\ns\frac{1}{\nicefrac{1}{4}\eta^2 - \xi^2}
\,,
\ee
is analytic on contour $\cC_{\infty}$, except at isolated points on the imaginary axis that can be avoided, and vanishes faster than 
$1/|\xi|$ for $|\xi|\rightarrow\infty$ which implies that the corresponding integral of the 
integrand in Eq.~\ref{eq-Tsuneto_omega=0-Matsubara} vanishes. For the small semi-circles $\xi=\pm\eta/2+\delta e^{i\theta}$ for $\theta\in\{0,\pi\}$. Integration around the small semi-circles yields,
\be
\int_{\cC_{\pm}}d\xi\,I(\xi)=
\mp\frac{i\pi}{\eta}\times\frac{1}{\varepsilon_n^2+|\Delta|^2+\nicefrac{1}{4}\eta^2}
\,.
\ee
Thus, $\int_{\cC_{+}+\cC_{-}}d\xi\,I(\xi) \equiv 0$, yielding a regular respsonse function for 
$\eta\rightarrow 0$ and $\fint d\xi\,I(\xi) = \ointctrclockwise_{\cC} d\xi\,I(\xi)$. Contour $\cC$ encloses a meromorphic integrand with a simple pole at $\xi=+i\sqrt{\varepsilon_n^2 + |\Delta|^2}$. Evaluating Eq.~\ref{eq-Tsuneto_omega=0-Matsubara} with the residue of the integrand yields Eq.~\ref{eq-Tsuneto_omega=0-regular}.

%
\end{document}